\documentclass[journal,10pt]{IEEEtran}
\usepackage{cite}
\usepackage{amsmath,amssymb,amsfonts}
\usepackage{caption}
\usepackage{subcaption}
\usepackage[ruled,vlined]{algorithm2e}
\usepackage{algpseudocode}
\usepackage{hyperref}
\hypersetup{colorlinks=true}
\usepackage{comment}
\algnewcommand{\Initialize}[1]{%
  \State \textbf{Initialize:}
  \Statex \hspace*{\algorithmicindent}\parbox[t]{.8\linewidth}{\raggedright #1}
}
\usepackage{graphicx}
\usepackage{color}
\DeclareMathOperator*{\argmin}{arg\,min}
\DeclareMathOperator*{\argmax}{arg\,max}
\newcommand{\PK}{}
\newcommand{\PS}{}
\usepackage{textcomp}
\def\BibTeX{{\rm B\kern-.05em{\sc i\kern-.025em b}\kern-.08em
    T\kern-.1667em\lower.7ex\hbox{E}\kern-.125emX}}
\begin{document}

\title{Memory Enabled Bumblebee-based Dynamic Spectrum Access for Platooning Environments}

\author{Kuldeep S. Gill,~\IEEEmembership{Student Member}, Pawel Kryszkiewicz,~\IEEEmembership{Senior Member}, Pawel Sroka,~\IEEEmembership{Member}, Adrian Kliks,~\IEEEmembership{Senior Member}, Alexander M. Wyglinski~\IEEEmembership{Senior Member}

\thanks{This work has been generously supported by the US National Science Foundation via Award Number 1547296 and the National Science Centre of Poland via Project Number 2018/29/B/ST7/01241.}
\thanks{Manuscript received October 1, 2021; revised November 30, 2022; accepted December 30, 2022.  final manuscript: 10.1109/TVT.2023.3236035}}

\makeatletter
\def\ps@IEEEtitlepagestyle{
  \def\@oddfoot{\mycopyrightnotice}
  \def\@evenfoot{}
}
\def\mycopyrightnotice{
  {\footnotesize
  \begin{minipage}{\textwidth}
  \centering
  Copyright~\copyright~2022 IEEE. Personal use of this material is permitted. However, permission to use this  \\ 
  material for any other purposes must be obtained from the IEEE by sending a request to pubs-permissions@ieee.org.
  \end{minipage}
  }
}

\markboth{IEEE Transactions on Vehicular Technology,~Vol.~XX, No.~XX, XXX~2021}
{}


\maketitle

\begin{abstract}
In this paper, we propose a novel memory-enabled non-uniform sampling-based bumblebee foraging algorithm (MEB) designed for optimal channel selection in a distributed Vehicular Dynamic Spectrum Access (VDSA) framework employed in a platoon operating environment. Given how bumblebee behavioral models are designed to support adaptation in complex and highly time-varying environments, these models can be employed by connected vehicles to enable their operation within a dynamically changing network topology and support their selection of optimal channels possessing low levels of congestion to achieve high throughput. As a result, the proposed VDSA-based optimal channel selection employs fundamental concepts from the bumblebee foraging model. In the proposed approach, the Channel Busy Ratio (CBR) of all channels is computed and stored in memory to be accessed by the MEB algorithm to make the necessary channel switching decisions. Two averaging techniques, Sliding Window Average (SWA) and Exponentially Weighted Moving Average (EWMA), are employed to leverage past samples and are evaluated against the no-memory case. Due to the high variability of the environment (\textit{e.g.}, high velocities, changing density of vehicles on the road), we propose to calculate the CBR by employing non-uniform channel sampling allocations as well as evaluate it using both simplified numerical and realistic Vehicle-to-Vehicle (V2V) computer simulations. The numerical simulation results show that gains in the probability of the best channel selection can be achieved relative to a uniform sampling allocation approach. By utilizing memory, we observe an additional increase in the channel selection performance. Similarly, we see an increase in the probability of successful reception when utilizing the bumblebee algorithm via a system-level simulator. 
\end{abstract}


\IEEEpeerreviewmaketitle

\section{Introduction}
\label{sec:introduction}
\IEEEPARstart{C}{onnected} vehicles leveraging Vehicle-to-Vehicle (V2V) communications will be able to exchange situational awareness information using Basic Safety Messages (BSM) to enhance driving experience and safety~\cite{park2011integrated}. As the demand for connected vehicles increases, existing spectrum allocation for V2V applications will be insufficient to meet future data traffic requirements~\cite{Shi2014}. Consequently, this will lead to frequency channel congestion in spectral bands allocated by the Federal Communication Commission (FCC) and other federal regulators, and will potentially increase the Packet Error Rate (PER) due to significant interference occurring between transmitting vehicles. While wireless networks employing Cellular-Vehicle-to-Everything (C-V2X) technology could potentially serve as a solution for enabling V2V connectivity, it also will eventually reach the same spectrum bottlenecks despite its advantages.

Vehicular Dynamic Spectrum Access (VDSA)~\cite{chen2011feasibility, pagadarai2009characterization} can yield a viable solution, where vehicles can leverage underutilized licensed spectrum to enable wireless access without interfering with the primary users of that spectrum. Within this VDSA framework, vehicles will be assigned as secondary users when accessing these licensed bands. Over the past decade, the FCC and other federal spectrum regulators around the world have opened up selected frequency bands for secondary access, such as Digital Television (DTV) spectrum, known as TV White Spaces (TVWS)~\cite{mishra2010much,kryszkiewicz2021dynamic}. In general, potentially other frequency bands can also be used for dynamic spectrum access. Regarding spectrum access technologies for V2V communications, currently the IEEE 802.11p Dedicated Short Range Communication (DSRC) standard~\cite{kenney2011dedicated} and the C-V2X standard~\cite{chen2020vision} are two frequently used solutions.{\PS Given the limited transmission range of intra-platoon communications, the technology that seems the best suited for VDSA is 802.11p as it can operate with lower channel access delays}, and can be deployed more quickly relative to C-V2X~\cite{8746562}. Moreover, its medium access protocol utilizes the ``Listen before talk'' mechanism, which makes V2V communications more robust against random interference that is more likely to occur in secondary spectrum access environments.  

VDSA has been recently investigated in various contexts. For example, authors in \cite{Liu2021} discuss the application of reinforcement learning algorithms (Q-Learning) for the cognitive internet of vehicles to improve utilization of spectrum and minimize the interference to primary users. In \cite{Arteaga2019}, the problem of the prospective lack of available spectrum for intelligent transportation systems was analyzed. In particular, the authors considered opportunistic access to DTV band, when - besides the television (TV) transmitters - other transmissions are also present, such as White-Fi networks. VDSA can potentially be advantageous in platooning scenarios, as it allows all intra-platoon communications to be switched to a single secondary channel, thus not requiring coordination with other vehicles or systems. Additionally, the inter-vehicle distances within a platoon is kept relatively small to minimize fuel consumption. Such an approach requires highly coordinated platoon control, \textit{e.g.}, Cooperative Adaptive Cruise Control (CACC)  for autonomous driving operations within the platoon, which relies on the continuous exchange of vehicle data. Thus, high quality inter-vehicles communications is needed, which may not be guaranteed if the communication bands are spectrally congested \cite{vukadinovic20183gpp, sybis2018context}. 
Consequently, VDSA can help achieve sufficient spectrum access ultimately supporting error-free CACC. Using VDSA, the platoon leader can select the optimal channel with the least amount of interference to achieve efficient V2V communications.

An example of a candidate frequency band to be considered for supporting CACC via V2V secondary transmission is the DTV spectrum~\cite{chen2011feasibility,kryszkiewicz2021dynamic}, which is mainly due to the temporal and spatial stability of the primary users within this band. In particular, the deployment of DTV primary users (TV towers, household TV receivers) as well as the frequency channel allocations for DTV towers vary slowly over time. In this scenario, the Radio Environment Map (REM) databases can be utilized in a centralized manner for finding the best spectrum opportunities \cite{Wei_2013}. REM databases may contain various kinds of information regarding the primary users (DTV transmissions), but they can also store the calculated allowed transmission power for V2V communications in the geographical area of interest~\cite{sybis2018context}. Another approach is to apply VDSA in a distributed manner, where individual vehicles or vehicular platoons discover the vacant channels by performing spectrum sensing. Therefore, radio environment awareness needs to be implemented in the platoon framework or even each vehicle to adapt to the time-varying channel conditions. 

A commonly used method for spectrum sensing is energy detection~\cite{urkowitz1967energy}, which possesses low implementation complexity and can quickly output results, which is critical for V2V communications. While the typical problem of energy detection is to obtain a sufficient sensitivity in the presence of noise, this is not the only limiting factor of a V2V communications environment. In fact, the primary challenge here is bursty interference. Persistent interference can be assessed via a single sweep of an energy detection process at each geographical location. However, in case of bursty interference, a single sweep of an energy detection scheme can result in the channel being detected as completely vacant or occupied at a single instant in time, but this may not be the case at another time instant. {\PS Such phenomena can be observed within the V2V communication environment, \textit{e.g.} using the IEEE 802.11p standard, which }employs bursty transmission that may subsequently generate bursty interference for other users. 

Once spectrum sensing has been performed, the platoon searches for the optimal frequency channel to use. The optimal frequency channel possesses the lowest Channel Busy Ratio (CBR). The CBR is also referred to by some authors as Channel Occupancy Ratio or Duty Cycle. It represents the probability that at a given time instance there will be an active transmission within the given channel. Alternatively, it can be defined as the ratio of time when the channel is busy versus the total observation time~\cite{Bazzi2019}. The challenge is how to reliably find the channel possessing the lowest CBR  across all considered channels using a limited number of samples during a rapidly varying operating environment resulting from vehicles movement. It should be noted that CBR estimation procedures have rarely been discussed in the literature. In \cite{Lehtomaki2015_CBR_estimation}, the authors focus on the optimization of CBR estimation for weak signals that are close to the noise floor. In particular, they considered the non-zero probabilities of false alarm and miss detection. Unfortunately, the discussion was limited to a single channel case rather than the multi-channel scenario considered in our research. Thus, in our work we concentrate on the multi-channel CBR estimation while applying non-uniform allocation of sampling instants across the channels. Additionally, we consider the effects of time-varying CBR, which occurs in realistic V2V environments where vehicles employ VDSA.

As mentioned previously, we are investigating the dynamic channel selection for V2V communications realized in a fully distributed manner, where each vehicle or platoon is making an individual channel decision. This approach is similar to bumblebee foraging behavior when selecting the best routes to collect pollen from flowers. In this paper, we propose a channel selection procedure utilizing a limited number of spectrum sampling results and a bumblebee behavioral model~\cite{gill2018memory} where the bumblebee foraging algorithm is a distributed optimization approach employing minimum social interaction requirements relative to the case of centralized optimization techniques~\cite{cheng2011cognitive, li2013adaptive}. To leverage the potential of bumblebee foraging behavior in connected vehicle environments, we translate the evolutionary optimized memory-mediated bumblebee foraging strategies to a VDSA decision-making algorithm used in connected vehicle networks. In~\cite{gill2018memory}, we demonstrated the potential utility of a bumblebee-inspired memory-based decision mechanism within a VDSA framework. The algorithm was evaluated using Geometry-Based Efficient Propagation Model Vehicles for V2V Communications (GEMV2)~\cite{boban2014geometry, gemv2} and highlighted the novelty of our proposed algorithm. The quantitative performance bound was simulated using multi-channel queuing theory for the bumblebee foraging algorithm~\cite{gill2018capacity}. In~\cite{gill2018experimental}, an experimental study was conducted for the bumblebee-inspired channel selection in an ad-hoc network using a collection of Pluto~\cite{pluto} Software-Defined Radio (SDR) units from Analog Devices. 

The novelty and the contributions of this work are the following:
\begin{itemize}
    \item Analytical bounds for the probability of selecting a channel with the lowest CBR is derived given the limited number of sensing samples per channel.
    \item Optimal sensing samples allocation strategies for \textit{a priori} known CBRs are derived.
    \item A heuristic approach for unequal samples distribution is provided for unknown CBRs, obtaining performance close to the optimal in a simulation environemnt.
    \item A bumblebee behavioral memory model is used to adapt the above mentioned heuristic to an environment of time varying CBRs and non-zero channel switching cost, and tested via simulation for Sliding Window Average (SWA) and Exponentially Weighted Moving Average (EWMA) memory models.
    \item A distributed VDSA channel sensing and selection algorithm is proposed employing the proposed heuristic unequal samples distribution and memory-based model for identification of the most promising channel.
    \item The proposed strategy is evaluated in sophisticated system-level simulations of motorway/highway platooning, comparing the equal and unequal sampling distribution strategies via showing probabilities of successful intra-platoon frame reception.
\end{itemize}

The rest of this paper is organised as follows:
Section~\ref{sec:system} presents the system model. In Section~\ref{sec:sampl}, a derivation of the upper and lower bounds of the probability for the best channel selection is presented in the case of fixed and known CBR values. This is followed by a numerical analysis and simulations showing how an unequal number of samples allocated can result in an improved probability for the best channel selection. While the optimal samples allocation is mathematically intractable for real-time platoon operation when the real CBR values are unknown, the unequal sampling heuristic approach is derived such that we obtain the quasi-optimal probability of best channel selection. The heuristic approach succeeds in finding the optimal channel faster for a broader range of system parameters. Next, in Section~\ref{sec:bumblebee} we expand upon the heuristic approach to a realistic time-varying CBR scenario where the memory-enabled bumblebee algorithm is implemented. {\PS The time-varying CBR model was acquired with the use of an advanced system-level V2V communications simulator \cite{sybis2019communication} and employed in the evaluation of the proposed VDSA solution.}
Two memory models are used
with the bumblebee algorithm to find the optimal channels in time. Their performance is compared via simulations in Section~\ref{sec:results} and recommendations for proper configuration are given. 
Finally, the proposed channel selection framework is evaluated in an advanced system-level simulator of VDSA for platooning in Section~\ref{sec:sls}. The uniform and non-uniform sampling scenarios{\PS , configured following the recommendations given in Sec. \ref{sec:results},} are compared, with results showing an improved reception rate with the heuristic approach due to the higher probability of best channel selection.
We conclude our work with closing remarks in Section~\ref{sec:conclusion}.

\section{System Model}
\label{sec:system}
\begin{figure}[th!]
    \centering
    \includegraphics[width=\columnwidth]{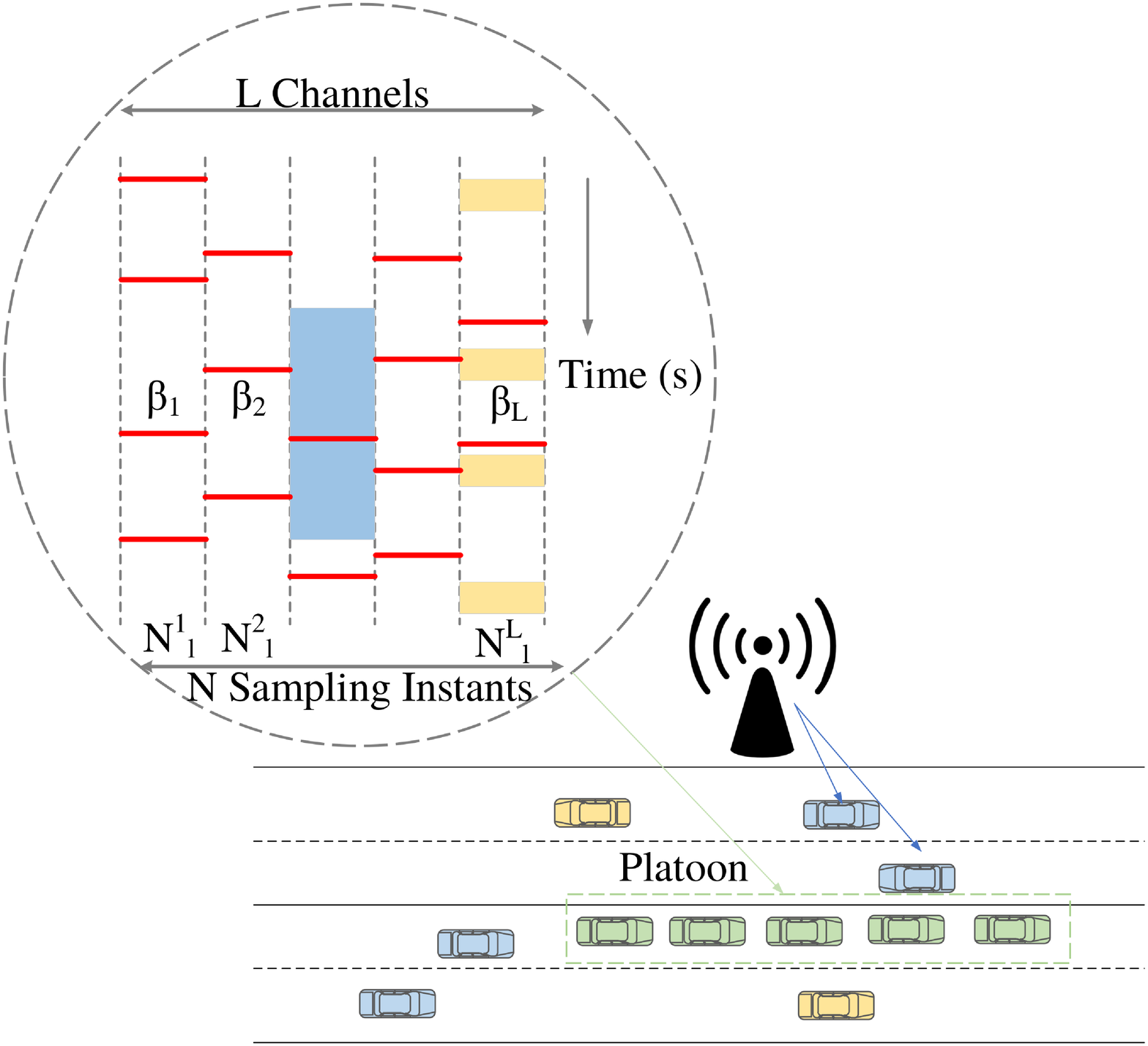}
    \caption{CBR $\beta$ variation with vehicles arriving and departing the transmission range of the platoon vehicles{\PS , with the first vehicle being the leader of the platoon}. Different channels have different $\beta$ values based on the channel utilization. Non-uniform channel sampling is performed by the platoon in order to find optimal channel from $L$ channels with the least $\beta$. There are total $N$ sampling periods which are distributed across the channels to optimize the sampling resources.}
    \label{fig:systemdiag}
\end{figure}

 We consider a VDSA system illustrated in Fig.~\ref{fig:systemdiag}, with high-speed road conditions consisting of four lanes, with two lanes devoted to each direction. The vehicles are exchanging {\PS messages using V2V communications, operating in a dynamically selected frequency band from a predefined set,} as it is assumed the nominal 5.9 GHz band is congested. {\PS Additionally, other stationary transmitters might be present in the system, such as the roadside infrastructure (represented as the black access point in Fig. \ref{fig:systemdiag}), thus being a source of additional interference in selected channels.\\ 
 The platoon of vehicles is marked with a dashed rectangle and is represented by the platoon leader, which is assumed to be the first vehicle in the platoon (the actual selection of platoon leader is out of scope of this work). The platoon leader will be responsible for selecting the least occupied channel. We assume that all platoon vehicles exchange information periodically, with the leader being responsible for the management of channel selection and the related reconfiguration procedure. While there are numerous information flow topologies proposed for platooning, the exact messaging scheme is outside the scope of this paper and a topic for a future publication by the authors. With every proposed communication protocol for platooning, the primary focus of this work is on maximizing the reliability of the communications, we try to achieve with the selection of the least occupied channel.\\
 All the devices (vehicles and infrastructure)} communicate over $L$ frequency channels indexed by $l= 1,...,L$. In Fig.~\ref{fig:systemdiag}, the bursty transmission of the yellow and blue vehicles is depicted via a dashed circle. The traffic generated by each vehicle is -- from the perspective of the platoon -- random and of various intensity and duration. {\PS Although V2V transmissions are generally periodic or quasi-periodic for a given service, such as the BSMs, the total generated traffic may be regarded as random by an external observer, due to the different configurations of the multiple services used by each vehicle.} {\PK Observe in each of $L$ considered frequency channels that different transmissions can occur, \textit{e.g.}, various frame duration, channel access scheme. While we assume the details of these transmissions are not known to the platoon, the probabilistic channels characterization is appropriate.} 
  Therefore, each channel is characterized by its Channel Busy Ratio (CBR) denoted as $\beta_l~\in~\langle 0,1\rangle$ for the $l$-th channel, which can be interpreted as ratio of time when the channel is busy to the total observation time. To detect the presence of the other ongoing transmissions non-uniform channel sampling is performed, with the sampling moments depicted as red lines.
 The goal of the considered platoon is to find the channel with the lowest CBR (denoted as $\hat{l}$) that allows for the highest reliability of intra-platoon transmission, \textit{i.e.}:
\begin{equation}
    \hat{l}=\argmin_l \beta_l.
\end{equation}
It is possible that more than one channel will have the minimal value of $\beta_l$. We denote this set of \emph{optimal} channel indices as $\mathbb{O}=\{ \hat{l} | \beta_{\hat{l}}=\min_{l} \beta_l \}$. {\PK All these channels are equally good to be used by the considered platoon for its intra-platoon communications, \textit{i.e.}, guarantee the lowest probability of message collisions.} The complementary set comprising of \emph{wrong} channel indices is denoted as $\mathbb{W}$ such that $\mathbb{C}=\mathbb{W} \cup \mathbb{O}=\{ 1,...,L\}$, where $\mathbb{C}$ is the set of all available channels.  

While $\beta_l$ is not known by the vehicles, it has to be estimated using sensing.
The vehicles are performing periodic spectrum sensing to find the most suitable channel for transmission. In our investigation, we chose an energy detection with power threshold based on the Carrier Sense Multiple Access with Collision Avoidance (CSMA/CA) approach, which is used for example in IEEE 802.11p \cite{IEEE80211}, to detect signals that can block intra-platoon communications. While the vehicles in the considered scenario can move with velocities of up to 130~kph, the impact of Doppler effect on the sensing process is negligible and not considered in this evaluation, as the maximum value of the frequency shift of approximately 1.4~kHz in the 5.9 GHz channel has no impact on the results of considered power measurements.

Various V2V communication protocols are able to operate in significant interference conditions, \textit{e.g.}, the required Signal to Interference and Noise Ratio (SINR) level for the 802.11p transmission is 2 dB for the most robust Modulation and Coding Scheme (MCS) which is MCS  $0$. Therefore, the platoon has to detect only the interference that is capable of blocking its transmission, \textit{i.e.}, significantly above the signal noise floor. As such, we can safely ignore the non-zero probability of false alarm important for weak signal detection \cite{Lehtomaki2015_CBR_estimation}. In the simplest approach, the platoon estimates the $\beta_l$ value, which is based on the available $N_l$ sensing decisions for the $l$-th channel. 
The Maximal Likelihood (ML) estimator of $\beta_l$ is given by:
\begin{equation}
    \hat{\beta}_l =\dfrac{k_l}{N_l},
    \label{eq:cbr}
\end{equation}
where $k_l\in \{0,...,N_l\}$ is the number of times the $l$-th channel has been sensed as busy over $N_l$ sampling moments.
 While this estimator is consistent, \textit{i.e.}, converges to real $\beta_l$ as $N_l$ increases to infinity, in practice its accuracy is limited. This can be both a result of the finite number of sensing opportunities before the channel is selected, and the limited number of sensing results within a given time period due to the technical limitations of the applied sensors (\textit{e.g.}, a sensor cannot sample more frequency channels as a result of high frequency-switching latency). 

Suppose we extend the above model to account for the increasing amount of platoon knowledge. This knowledge is improved during each consecutive iteration, \textit{i.e.}, a selected time period allowing for intra-platoon communications, when performing spectrum sensing and selection of a new frequency channel to be used for communications. Each iteration, indexed with $(i)$, consists of a limited total number of $N$ sensing periods that can be carried over all $L$ frequency channels. 
At the $i$-th iteration the platoon has to decide on the number of sensing moments $N_l^{(i)}$ assigned to each channel $l\in\{1,...L\}$ so that the total limit of $N$ is reached, \textit{i.e.}, $\sum_{l=1}^{L}N_l^{(i)}=N$. Observe that we assume the total number of sensing samples $N$ is fixed, \textit{e.g.}, as a result of the total number of vehicles in the platoon and the fixed iteration duration.
Without loss of generality, it can be assumed the first iteration is $i=1$. 
Based on the number of times channel $l$ is detected as busy at iteration $i$, \textit{i.e.}, $k_l^{(i)}$, as well as the prior knowledge obtained from sensing in previous iterations, the estimate of $\beta_l$ is obtained. 
Assuming a stationary $\beta_l$, the ML estimator is specified as:
\begin{equation}
\label{eq_beta_est}
    \hat{\beta}_{l}^{(i)}=\frac{\sum_{j\leq i}k_l^{(j)}}{\sum_{j\leq i}N_l^{(j)}}= \frac{\tilde{k}_l^{(i)}}{\tilde{N}_l^{(i)}},
\end{equation}
where $\tilde{k}_l^{(i)}$ denotes the cumulative number of times the signal was detected, and $\tilde{N}_l^{(i)}$ is the cumulative number of samplings performed over all past iterations as well as the current $i$-th iteration. 

The platoon switches its carrier frequency to the one of the lowest expected occupancy, \textit{i.e.}:
\begin{equation}
\label{eq_channel_selection}
    \hat{l}=\arg \min_l \hat{\beta}_{l}^{(i)}.
\end{equation}
The question is how to distribute $N$ sampling moments among $L$ channels in such a way that the probability of the optimal channel selection is maximized in each iteration. The discussion on the optimal channel selection in the case of varying CBR values will be continued in Section~\ref{sec:bumblebee}.\\
{\PS 
Note that without loss of generality, the proposed approach can be extended to a multi-platoon scenario, where each platoon would treat the transmissions of other groups of vehicles simply as interference, influencing thus the estimated CBR values.
}

\section{Samples-to-channel allocation strategies} 
\label{sec:sampl}

\label{sec_samples_alloc}
The simplest approach is to assign an equal number of sensing moments (to be referred to from now on as \emph{equal allocation}) to each channel, \textit{i.e.}:
\begin{equation}
    N_l^{(i)}\approx \frac{N}{L}.
    \label{eq_simple_allocation}
\end{equation}
The approximation sign is used as the number $N_l^{(i)}$ has to be natural while the division of $N$ over $L$ can result with a non-integer value. In such a case, each channel will be assigned $\lfloor \frac{N}{L}\rfloor$ samples and the remaining $N-L\lfloor \frac{N}{L}\rfloor$ \footnote{The operation $\lfloor \cdot \rfloor$ represents the floor function, which outputs the greatest integer not greater than the input argument.} samples will be distributed randomly across all $L$ channels. However, after several iterations if the estimated $\beta_l$ for a set of channels is relatively low and very high for several others, the equal allocation would be not optimal. There is no point in increasing the accuracy of the CBR estimates for very busy channels as these have low probability of being selected as the optimal for V2V communications. The sampling should then be increased for the channels with relatively low CBR values as these are the ones that may contain the optimal channel(s) for data transmission, \textit{i.e.}, the channels with indices belonging to the set~$\mathbb{O}$. 

\subsection{Optimal Channel Sampling for \textit{A Priori} Known CBR Values}
This subsection focuses on the optimal allocation of samples among channels. The upper bound of the probability of the optimal channel selection is calculated while adjusting samples allocation assuming the real $\beta_l$ values are known. In order to do so, first an analytical formula for a tight Upper Bound (UB) and Lower Bound (LB) on a probability of success will be derived.

\subsubsection{Bounds of probability of success}
The probability of success, \textit{i.e.}, the selection of an optimal channel, can be calculated for a given number of samples allocated until the $i$-th iteration to channel $l$, denoted as $\tilde{N}_l^{(i)}$. 
Success is a random event that can be formally defined as $\hat{l}$ chosen using (\ref{eq_channel_selection}) belonging to $\mathbb{O}$.
Randomness is a result of $\hat{\beta}^{(i)}_l$ being a {\PK discrete random }variable that depends on the random values of $k_l^{(j)}$ and allocated $N_l^{(j)}$ for $j\leq i$ according to (\ref{eq_beta_est}). 
 The success will occur every time the minimal estimated CBR value of any optimal channel, \textit{i.e.}, $\hat{\beta}_{l}^{(i)}$ for $l\in \mathbb{O}$, will be smaller than minimal estimated CBR for any wrong channel, \textit{i.e.}, $\hat{\beta}_{l}^{(i)}$ for $l\in \mathbb{W}$. This is be defined as:
\begin{equation}
    \Pr\left\{ \min_{l\in \mathbb{O}} \hat{\beta}_{l}^{(i)}  <
    \min_{l\in \mathbb{W}} \hat{\beta}_{l}^{(i)} 
    \right\}\leq \Pr\left\{
    \hat{l} \in \mathbb{O}
    \right\},
    \label{eq_lower_bound}
\end{equation}
where the left hand side constitutes a LB, and the right hand side forms the probability of success.
The above defined probability provides the lower bounds for the total probability of success as it considers that for all the events  when $\min_{l\in \mathbb{O}} \hat{\beta}_{l}^{(i)}=
\min_{l\in \mathbb{W}} \hat{\beta}_{l}^{(i)} $ the wrong channel is selected. On the other hand, the upper bound of the probability of success can be derived counting all the events when $\min_{l\in \mathbb{O}} \hat{\beta}_{l}^{(i)}=
\min_{l\in \mathbb{W}} \hat{\beta}_{l}^{(i)} $ as success. This results in:
\begin{equation}
    \Pr\left\{
    \hat{l} \in \mathbb{O}
    \right\} \leq
    \Pr\left\{ \min_{l\in \mathbb{O}} \hat{\beta}_{l}^{(i)}  \leq
    \min_{l\in \mathbb{W}} \hat{\beta}_{l}^{(i)} 
    \right\}.
    \label{eq_upper_bound}
\end{equation}
It is observed the right-hand side of the above formula can be partitioned into a sum of probabilities, where the minimum estimated CBR for the optimal channels will be lower than the minimum for the wrong channels, and that both these minima will be equal, \textit{i.e.}:
\begin{align}
\label{eq_upper_bound_partitioned}
    &\Pr\left\{ \min_{l\in \mathbb{O}} \hat{\beta}_{l}^{(i)}  \leq
    \min_{l\in \mathbb{W}} \hat{\beta}_{l}^{(i)} 
    \right\}=
    \\\nonumber
    &\Pr\left\{ \min_{l\in \mathbb{O}} \hat{\beta}_{l}^{(i)}  <
    \min_{l\in \mathbb{W}} \hat{\beta}_{l}^{(i)} 
    \right\}+
    \Pr\left\{ \min_{l\in \mathbb{O}} \hat{\beta}_{l}^{(i)} =
    \min_{l\in \mathbb{W}} \hat{\beta}_{l}^{(i)} 
    \right\}.
\end{align}

Let us define the random variables
$b^{(i)} =\min_{l\in \mathbb{O}} \hat{\beta}_{l}^{(i)}$ and $c^{(i)} =\min_{l\in \mathbb{W}} \hat{\beta}_{l}^{(i)}$. 
The Cumulative Density Functions (CDFs) of $b^{(i)}$ and $c^{(i)}$ can be calculated assuming that $\hat{\beta}_{l}^{(i)}$ values are independent among channels. Additionally, for a given $\hat{\beta}_l^{(i)}$ and $N_l^{(i)}$ the number of times the signal is detected in channel $k_l^{(i)}$ should be independent from $k_m^{(i)}$ for $m\neq l$, \textit{e.g.}, as a result of adjacent channels not overlapping in frequency. 
The CDF of $b^{(i)}$ is denoted as
$F_{b^{(i)}}(x)=\Pr\left\{{b^{(i)}} \leq x \right\}$.
It is known that $1-\Pr\left\{{b^{(i)}} \leq x \right\}$
equals to $Pr\left\{{b^{(i)}} > x \right\}$, \textit{i.e.}, probability that all $\forall_{l\in \mathbb{O}} \hat{\beta}_{l}^{(i)}$ are greater than $x$. Utilizing the independence of the $\hat{\beta}_{l}^{(i)}$ random variables with CDF denoted as $F_{\hat{\beta}_{l}^{(i)}}(x)$ we obtain:
\begin{align}
    F_{b^{(i)}}(x)&=1-\Pr\left\{{b^{(i)}} > x \right\}
    =1-\prod_{l \in \mathbb{O}} \Pr\left\{\hat{\beta}_{l}^{(i)}  > x \right\}
    \nonumber \\ 
    &=1-\prod_{l \in \mathbb{O}} \left( 1-
    F_{\hat{\beta}_{l}^{(i)}}(x)
    \right) .
\end{align}
Following the same reasoning, the CDF for $c^{(i)}$ can be derived, yielding the expression:
\begin{align}
    F_{c^{(i)}}(x)=1-\prod_{l \in \mathbb{W}} \left( 1-
    F_{\hat{\beta}_{l}^{(i)}}(x)
    \right) .
\end{align}
According to (\ref{eq_beta_est}) the random variable $\hat{\beta}_{l}^{(i)}$ is a sum of binomial distributed random variables $k_l^{(j)}$ scaled by a given value $\tilde{N}_l^{(i)}$. Its CDF can be defined as:
\begin{equation}
\label{eq_CDF_beta}
    F_{\hat{\beta}_{l}^{(i)}}(x)=\sum_{q=0}^{\left\lfloor \min\left(
    1, x\right)\tilde{N}_l^{(i)} \right\rfloor}
   \binom{\tilde{N}_l^{(i)}}{q}\beta_l^q(1-\beta_l)^{
   \tilde{N}_l^{(i)}-q}.
\end{equation}
Recall that $F_{\hat{\beta}_{l}^{(i)}}(x)$ is 0 for $x<0$ and 1 for $x>1$.

Knowing $F_{c^{(i)}}(x)$ and $F_{b^{(i)}}(x)$, the lower bound defined in (\ref{eq_lower_bound}) can be calculated, yielding the probability of all events when the random variable $c^{(i)}$ is greater than $b^{(i)}$. The probability that a given random variable is greater than a given value is defined by a complementary cumulative density function, \textit{i.e.}, $1-F_{c^{(i)}}(x)$ for $c^{(i)}$. 
By the Law of the Total Probability, the probability of $c^{(i)}>x$ has to be added for all possible $x$ values being values of a random variable $b^{(i)}$ (for $c^{(i)}>b^{(i)}$), weighted by probabilities of $b^{(i)}$, \textit{i.e.}:
\begin{equation}
    \Pr\left\{ b^{(i)}  <c^{(i)}\right\}
    =\int_{0}^{1} f_{b^{(i)}}(x) \left( 1-F_{c^{(i)}}(x)\right) dx,
\end{equation}
where $f_{b^{(i)}}(x)$ is the Probability Density Function (PDF) of $b^{(i)}$. It is calculated using the differentiation operation:
\begin{equation}
    f_{b^{(i)}}(x)=\frac{d F_{b^{(i)}}(x)}{d x}=\sum_{l \in \mathbb{O} } \frac{d F_{\hat{\beta}_{l}^{(i)}}(x)}{d x}
	    \prod_{\hat{l} \in \mathbb{O}-l} 
	    \left( 1-
    F_{\hat{\beta}_{\hat{l}}^{(i)}}(x)\right),
    \label{eq_PDF}
\end{equation}
where: 
\begin{equation}
    \frac{d F_{\hat{\beta}_{l}^{(i)}}(x)}{d x}= \sum_{q=0}^{\tilde{N}_l^{(i)}}
   \binom{\tilde{N}_l^{(i)}}{q}
  \beta_l^q(1-\beta_l)^{\tilde{N}_l^{(i)}-q}
   \delta\left(x\tilde{N}_l^{(i)}-q \right), 
\end{equation}
and $\delta(\cdot)$ is Dirac delta function.
Observe that both $b^{(i)}$ and $c^{(i)}$ are specified in the range $\langle 0, 1\rangle$ that is reflected by the range of integral. This probability can be calculated accurately, {\PK \textit{i.e.}, without numerical approximation,} as all these distributions are for discrete variables. {\PK Observe that both $b^{(i)}$ and $c^{(i)}$ come from $\hat{\beta}_{l}^{(i)}$ being a discrete variable as shown by (\ref{eq_CDF_beta}). The integration can be replaced with the summation in a software implementation}.   

While the above formula gives the lower bound of the probability of success, the upper bound additionally requires the probability that minima of $\hat{\beta}_{l}^{(i)}$  for both the optimal and wrong channel sets are equal, according to (\ref{eq_upper_bound_partitioned}). This can be obtained by using the PDFs of $b^{(i)}$ and $c^{(i)}$, and, thanks to these events independence, integrate over all their possible values as:
\begin{equation}
    \Pr\left\{ b^{(i)} =
    c^{(i)} 
    \right\}= \int_{0}^{1} \int_{x=y} f_{b^{(i)}}(x)f_{c^{(i)}}(y)dx dy .
    \label{eq_prob_equal}
\end{equation}
The PDF $f_{c^{(i)}}(x)$ is calculated from $F_{c^{(i)}}(x)$ in a manner similar to the variable $b^{(i)}$ in (\ref{eq_PDF}).

The upper bound considers as a success for all the events when: 1) the minimal 
estimated CBR for the optimal channel is smaller than the minimal estimated CBR for a wrong channel, and 2) the minimal estimated beta values for both the optimal and wrong channel sets are equal. To tighten the upper bound, we focus on event 2). If this happens, the number of optimal channels having the minimum estimated CBR value ranges from 1 to $|\mathbb{O}|$, where $|\cdot|$ denotes the cardinality of a set. This number can be denoted as a random variable $q_{O}\in \{1, ..., |\mathbb{O}|\}$. Under the same condition, the number of wrong channels having the minimum estimated CBR value ranges from 1 to $|\mathbb{W}|$ and is denoted as a random variable $q_{W}$.
We can assume that the selection of each of these $q_{W}+q_{O}$ channels is equally probable while using (\ref{eq_channel_selection}). As such, the conditional probability of success, \textit{i.e.}, considering the event 2) happens, equals $q_{O}/(q_{O}+q_{W})$.
Unfortunately, the probability derived in (\ref{eq_prob_equal}) does not differentiate between various values of $q_{O}$ and $q_{W}$. For each of these $|\mathbb{W}||\mathbb{O}|$ events possibly different probability is expected. However, the upper bound can be defined assuming the maximal value of $q_{O}/(q_{O}+q_{W})$ factor for a non-empty set $\mathbb{W}$, \textit{i.e.}, $|\mathbb{O}|/(|\mathbb{O}|+1)$.
The tightened upper bound in respect to (\ref{eq_upper_bound}) can be defined as:
\begin{equation}
    \Pr\left\{
    \hat{l} \in \mathbb{O}
    \right\} \leq
    \Pr\left\{ b^{(i)} < c^{(i)}\right\}
    +\frac{|\mathbb{O}|}{|\mathbb{O}|+1} \Pr\left\{ b^{(i)} = c^{(i)}\right\}.
    \label{eq_upper_bound_impr}
\end{equation}
With the same reasoning, the lower bound (\ref{eq_lower_bound}) can be tightened, creating a new lower bound by considering not only all the events when $b^{(i)} < c^{(i)}$ but also some part of the events when  $b^{(i)} = c^{(i)}$. It can be observed that for all the events when $b^{(i)} = c^{(i)}$ the minimal value of factor $q_{O}/(q_{O}+q_{W})$ is $1/(|\mathbb{W}|+1)$. Therefore, the improved lower bound is:
\begin{equation}
    \Pr\left\{\! b^{(i)} \!<\! c^{(i)}\!\right\}
    +\frac{1}{|\mathbb{W}|\!+\!1} Pr\left\{ b^{(i)} \!=\! c^{(i)}\!\right\}
    \!\leq\!
    \Pr\left\{\!
    \hat{l} \in \mathbb{O}
    \!\right\}  .
    \label{eq_lower_bound_impr}
\end{equation}
Finally, the lower and upper bounds to be used are defined by (\ref{eq_lower_bound_impr}), and (\ref{eq_upper_bound_impr}), respectively.

\subsubsection{Probability of success for optimal channel sampling}
\label{sec_optimal_prob}
The previous section provided mathematical derivations for the tight bounds for the probability of success in channel selection, knowing the samples allocation for each channel $\tilde{N}_l^{(i)}$ as defined in (\ref{eq_beta_est}) and CBR $\beta_l$. Now, the values of $\tilde{N}_l^{(i)}$ can be optimized to obtain the maximum possible probability of success. As the upper bound is being assessed, the usage of the upper bound derived in the previous section is justified. 

\noindent\underline{Global optimal solution:}
Let us first find global optimal solutions, \textit{i.e.}, what should be the total number of samples taken in channel $l$ over the current and all the previous iterations $j\leq i$, \textit{i.e.}, $\tilde{N}_l^{(i)}$, to maximize the probability of success in the $i$-th iteration. Every possible solution to this problem can be represented as a $L$-tuple $\mathbf{n}^{(i)}=\left(\tilde{N}_1^{(i)},...,\tilde{N}_L^{(i)}\right)$ for which each element is a natural number, and the number of all samples taken over all channels is $iN$, \textit{i.e.}, $\sum_{l=1}^{L} \tilde{N}_l^{(i)}=iN$. In general, the total number of such combinations is equal to $\binom{L-1+iN}{L-1}$. However, before any knowledge about CBR of any of channels is obtained, it is reasonable to assume that in the first iteration  the \emph{equal allocation} approach is used. As such, $\lfloor \frac{N}{L}\rfloor$ samples constitute a fixed allocation of samples per each element of each possible tuple that is possibly increased by any combination 
of $iN-L\lfloor \frac{N}{L}\rfloor$ samples distribution among $L$ channels. There are $\binom{L-1+iN-L\lfloor \frac{N}{L}\rfloor}{L-1}$ possible allocations. These can be generated, \textit{e.g.}, using \emph{stars and bars} method, and will be denoted as a set $\mathbb{N}^{(i)}_{\mathrm{global}}$. While the total number of possibilities increases quickly with the iteration index $i$ and the number of samples allocated per iteration $N$, the computational complexity of finding the global optimal solution can be significant.   

The optimization problem is defined as:
\begin{equation}
    \mathbf{\hat{n}^{(i)}}=\argmax_{
    \mathbf{n}^{(i)} \in
    \mathbb{N}^{(i)}_{\mathrm{global}}
    } 
    \Pr\left\{ b^{(i)} < c^{(i)}\right\}
    +\frac{|\mathbb{O}|}{|\mathbb{O}|+1} \Pr\left\{ b^{(i)} = c^{(i)}\right\},
\end{equation}
where $\mathbf{\hat{n}}^{(i)}$ is the optimal tuple of samples taken in each channel in $i$-th iteration.
It is solved by exhaustive search of the solution space, independently for each iteration $i$.

\noindent\underline{Iterative optimal solution:}
While the global method maximizes the probability of optimal channel selection in $i$-th iteration, its samples allocation does not consider how the samples were allocated in the previous iterations or what was the previous probability of success. As such, this upper bound can result, \textit{e.g.}, in the optimal cumulative number of samples allocated to decrease over consecutive iterations, \textit{i.e.}, $\tilde{N}_l^{(j)}>\tilde{N}_l^{(i)}$ for $j<i$. This is not possible in a practical sensing system when in each iteration a decision on samples distribution is made and cannot be changed (the historical allocation) in the next iteration. In each iteration, only $N$ samples can be allocated among $L$ channels.  Therefore, in $i$-th iteration ($i>1$), the solution space is defined as a set $\mathbb{N}^{(i)}_{\mathrm{iter}}$ of $\binom{L-1+N}{L-1}$ $L$-tuples $\mathbf{n}^{(i)}_{\mathrm{iter}}$. Each element of this tuple is defined as $\hat{N}_l^{(i-1)}+{N}_l^{(i)}$ where $\hat{N}_l^{(i-1)}$ is the cumulative samples allocation $\tilde{N}_l^{(i-1)}$ over all previous iterations, that has been decided as a result of optimization in the previous iterations. The chosen allocation from the $i$-th iteration over all channels, composed of elements $\hat{N}_l^{(i)}$, is denoted as a tuple $\mathbf{\hat{n}}^{(i)}_{\mathrm{iter}}$. 

In the case of the first iteration, $\lfloor \frac{N}{L}\rfloor$ samples are initially allocated to each channel. As such, only $N-L\lfloor \frac{N}{L}\rfloor$ samples have to be optimally allocated among $L$ channels. In this case, the solution space $\mathbb{N}^{(1)}_{\mathrm{iter}}$ is composed of $\binom{L-1+N-L\lfloor \frac{N}{L}\rfloor}{L-1}$ tuples. 

The optimization problem is defined as:
\begin{equation}
   \mathbf{\hat{n}}^{(i)}_{\mathrm{iter}}=\argmax_{
    \mathbf{n}^{(i)}_{\mathrm{iter}} \in
    \mathbb{N}^{(i)}_{\mathrm{iter}}
    } 
    \Pr\left\{ b^{(i)}\! <\! c^{(i)}\right\}
    +\frac{|\mathbb{O}|}{|\mathbb{O}|+1} \!\Pr\left\{ b^{(i)} \!=\! c^{(i)}\!\right\}.
\end{equation}
Most importantly, in this case the optimization has to be carried iteratively, each time increasing $i$ by $1$ as the solution space $\mathbb{N}^{(i)}_{\mathrm{iter}}$ depends on $ \mathbf{\hat{n}}^{(i-1)}_{\mathrm{iter}}$, being the resultant samples allocation in the previous iteration. The computational complexity of solving this problem is moderate as the solution space $\mathbb{N}^{(i)}_{\mathrm{iter}}$ does not scale with the iteration index $i$, but depends only on the total number of samples assigned per one iteration $N$, assuming a fixed number of observed channels $L$. 

While implementing the iterative solutions, it is important to consider there might be several optimal solutions, \textit{i.e.}, having the same upper bound of probability of success in the $i$-th iteration. However, depending on the choice in the $i$-th iteration, the achievable upper bound can vary in subsequent iterations. The suggested solution is to evaluate in parallel all the equally optimal solutions until several of them become better relative to the others in the following iterations.  

\subsection {Analysis of Optimal Allocation and Proposal of Unequal Samples Allocation}
\label{sec_proposal}
\begin{figure}[!ht]
    \centering
    \includegraphics[width=0.45\textwidth]{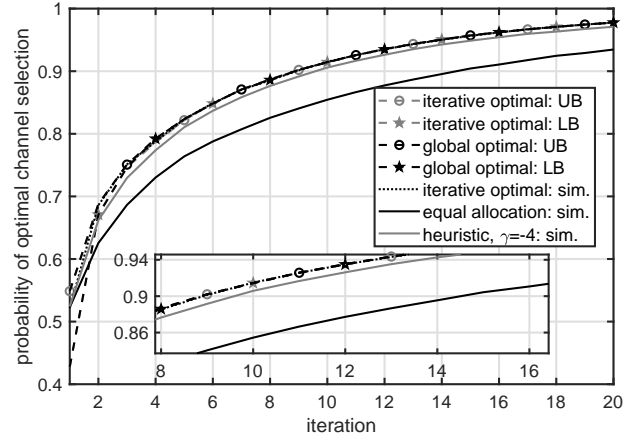}
    \caption{Probability of optimal channel selection versus iteration for $\beta_1=0.2$, $\beta_2=0.35$, $\beta_3=0.6$, $\beta_4=0.8$ and $N=8$. }
    \label{fig:prob_N8}
\end{figure}

To observe how the optimal allocation differs from an equal allocation, computer simulations were performed. The following arbitrary parameters have been selected: four channels ($L=4$) with $\beta_1=0.2$, $\beta_2=0.35$, $\beta_3=0.6$ and $\beta_4=0.8$. 
The optimal samples distribution were derived using both iterative and global methods presented in Section~\ref{sec_optimal_prob}. Both upper and lower bounds for both optimal sampling methods are shown in Fig.~\ref{fig:prob_N8} for $N=8$, which shows that both UB and LB are very close to each other, and visibly overlap after the third iteration. As expected, the probability of optimal channel selection increases with a growing number of iterations. In addition to the analytical results, the simulation results are presented, where in each case $10^5$ independent, random runs are performed to obtain sufficient statistical correctness. First, an \emph{iterative optimal: sim.} series is obtained for samples allocated according to \emph{iterative optimal} method. The simulation result is bounded by the iterative optimal UB and LB that can be treated as a partial confirmation of the correctness of the derived bounds. However, this optimal samples allocation approach is obtained knowing $\beta_i$ values. Without this knowledge, a standard solution is \emph{equal allocation}, meaning there are two new samples (for the considered case, when $N=8$ and $L=4$) obtained for each channel in each iteration. The equal allocation is visibly worse than an optimal allocation, yielding the probability of the best channel selection to be 0.9 after nearly 6 iterations relative to such a success probability is achieved for the optimal samples allocation. Equivalently, the 0.9 probability of optimal channel selection will be achieved taking 60\% more time for the equal allocation algorithm.
\begin{figure}[htb]
    \centering
    \includegraphics[width=0.45\textwidth]{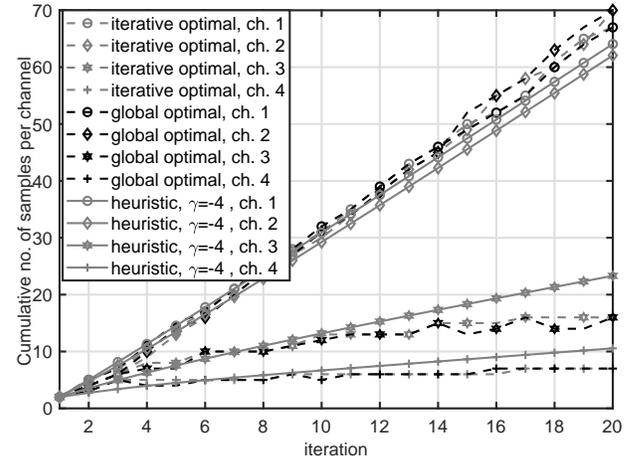}
    \caption{Cumulative number of sampling moments per channel (ch.) versus iteration for $\beta_1=0.2$, $\beta_2=0.35$, $\beta_3=0.6$, $\beta_4=0.8$ and $N=8$. }
    \label{fig:No_samples_N8}
\end{figure}

Let us analyze the cumulative number of samples per channel taken for the optimal allocation schemes given the above defined set of parameters in Fig.~\ref{fig:No_samples_N8}. 
We observe that it rises nearly linearly for each channel for both iterative and global optimal methods. However, the most significant increase in the number of samples is observed for the optimal channel (channel 1) and the wrong channel of the lowest CBR (channel 2). After 20 iterations, each of these channels was sampled nearly 70 times (approximately 3.4 samples per iteration out of 8 available). At the same time, the third and fourth channels are sampled with approximately 0.8 and 0.35 samples per iteration, respectively. 
This is the motivation for a definition of the unequal samples allocation heuristic approach:
\begin{equation}
    N_l^{(i)}\approx \frac{n_l^{(i)}}
    {\sum_{q=1}^{L}n_q^{(i)}}N,
    \label{eq_unequal}
\end{equation}
where:
\begin{equation}
    n_l^{(i)}=
    \begin{cases}
    \exp \left(\gamma\hat{\beta}_{\tilde{l}}^{(i-1)}\right) &\text{for } l=\hat{l}\\
    \exp \left(\gamma\hat{\beta}_{l}^{(i-1)}\right)  &\text{elsewhere},
    \end{cases}
\label{eq_heuristic}
\end{equation}
$\hat{l}$ is obtained according to (\ref{eq_channel_selection}), $\tilde{l}$ is the channel index of the second-best estimated channel, \textit{i.e.}, $\tilde{l}=\arg \min_{l\in\{1,...,\hat{l}-1,\hat{l}+1,...,L\} } \hat{\beta}_{l}^{(i-1)}$ and $\gamma$ is {\PK an empirical parameter for finding balance between equal and unequal samples allocation}. The approximation in (\ref{eq_unequal}) is required as the right-hand side values are not necessarily integers, and can require rounding to obtain $N_l^{(i)}$ summing up to $N$. The expected range for $\gamma$ is from $-\infty$ to $0$. While for $\gamma=0$ an equal allocation is obtained, the lower the $\gamma$ value, the higher portion of samples is expected to be assigned to the channels of relatively low estimated CBR. 

As an initial test this heuristic approach is simulated for the system configuration mentioned previously with an arbitrarily chosen $\gamma=-4$. The results are depicted in Fig.~\ref{fig:prob_N8} and Fig.~\ref{fig:No_samples_N8}. The probability of the best channel selection is very close to the derived upper bound, delayed by approximately half of an iteration at the 0.9 probability level. At the same time, the cumulative number of samples per channel averaged over all simulation runs resembles the plots for global optimal solution.

\begin{figure}[!ht]
    \centering
    \includegraphics[width=0.45\textwidth]{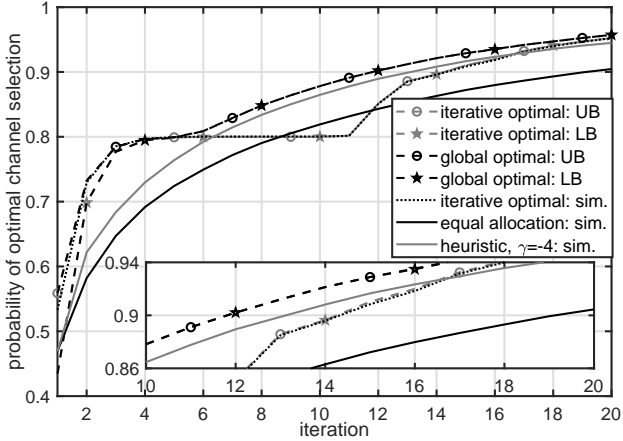}
    \caption{Probability of optimal channel selection versus iteration for $\beta_1=0.2$, $\beta_2=0.35$, $\beta_3=0.6$, $\beta_4=0.8$ and $N=6$. }
    \label{fig:prob_N6}
\end{figure}
To show the proposed scheme works well even for sparsely sensed channels, the number of samples available per each iteration is reduced to $N=6$. The resultant probability of the best channel selection as a function of the number of iterations is shown in Fig.~\ref{fig:prob_N6}. 
First, let us focus on the global and iterative UB and LB. The upper bounds overlap similarly as lower bounds until sixth iteration. The gap between the UB and LB is the largest for the first iteration and equals about 0.13, but it  rapidly decreases as the number of iterations increases. One interesting observation is the separation of iterative optimal bounds and global optimal bounds between sixth iteration up to the twentieth iteration. While both optimal solutions achieve the probability of the best channel selection equaling 0.8 relatively quickly, the iterative solutions reach a plateau at this level. This behavior is precisely reflected by simulations, \textit{i.e.}, ``iterative optimal: sim.'' series. Most interestingly, around the tenth iteration the equal allocation strategy outperforms the iterative optimal solution. The greedy solution, \textit{i.e.}, the maximization of the probability of the best channel selection for each iteration, results in a temporal deadlock. This brings us to the conclusion that there cannot exist a samples allocation strategy that always achieves the global UB in all iterations for all possible system configurations. 

From this perspective, the proposed heuristic will perform well close to the global optimal solution for higher number of iterations. While the best channel is selected with probability 0.9 after 12 iterations for the global optimal solution, the system utilizing the proposed heuristic requires one more iteration. The same reliability is achieved with the equal samples allocation after 19 iterations. 

\begin{figure}[htb]
    \centering
    \includegraphics[width=0.45\textwidth]{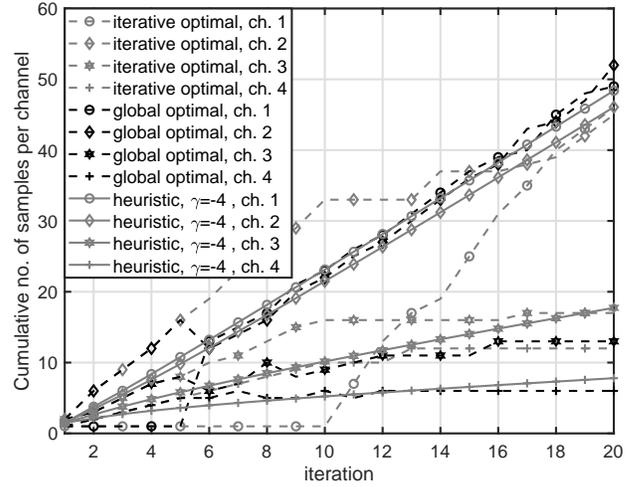}
    \caption{Cumulative number of sampling moments per channel versus iteration for $\beta_1=0.2$, $\beta_2=0.35$, $\beta_3=0.6$, $\beta_4=0.8$ and $N=6$. }
    \label{fig:No_samples_N6}
\end{figure}
A justification with respect to the plateau occurring in the probability of the best channel selection can be found by observing the cumulative number of sampling moments per channel, which is given in Figure~\ref{fig:No_samples_N6}. Until the tenth iteration in the iterative optimal solution, channel~1 is sampled only with a single sample. In 80\%  of cases, this results in the estimated CBR, \textit{i.e.}, $\hat{\beta}^{(i)}_1$, being equal to 0 as $\beta_1=0.2$. On the other hand, the remaining heavily sampled channels can achieve a CBR of 0 with low probability, with the optimal channel successfully selected in around 80\% of cases. However, to achieve the probability of optimal channel selection being higher than 0.8, the first channel sensing scheme has to be significantly changed. Between the fifth and sixth iteration for global optimal solution, we observe the cumulative number of samples for the first channel rising from 1 to 13 while at the same time the number of samples drawn from the second channel falls from 16 to 12. While this is optimal globally, it cannot be achieved with an iterative solution that cannot decrease the number of cumulative past samples per channel while proceeding with iterations. After the tenth iteration, the first channel starts to be sampled the most frequently in the case of an iterative optimal solution (observed having the highest gradient of values).
Most importantly, the proposed heuristic approach sufficiently approximates the globally optimal solution after the sixth iteration, which suggests it can be used to improve the probability of optimal channel selection even for a limited number of samples per iteration~$N$. 

\begin{figure}[htb]
    \centering
    \includegraphics[width=0.45\textwidth]{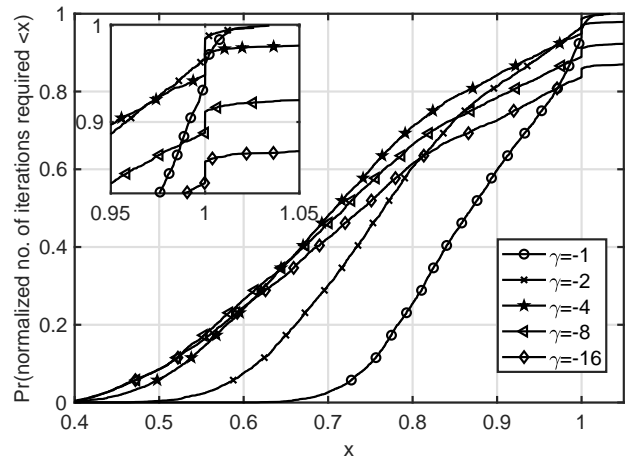}
    \caption{CDF of normalized number of iterations required to obtain probability of best channel selection equal 0.95 over $5,200$ random system configurations (varying $N$, $L$, and $\beta_i$). Normalization over number of iterations required by equal allocation.}
    \label{fig:CDF_gamma_change}
\end{figure}

As previously discussed, the proposed heuristic approach works well for two considered system configurations, \textit{i.e.}, sets of $\beta_i$, $N$, and $L$ values. However, the consistency of this behavior should be verified for other system configurations. Additionally, a recommendation regarding the choice of $\gamma$ values should be provided. 
As a result, a set of 26 different $N$ and $L$ values was studied. The focus was on systems where the numbers of available samples per iteration is small as this is the most challenging scenario for platooning using VDSA. For each of these $L$ and $N$ configurations, $200$ random $\beta_i$ sets (each of length $L$) were chosen with $\beta_i\in \{0, 0.1,...,1\}$. The limited set of possible CBR values mimics the discrete character of the CBR introduced by each vehicle transmitting in a given channel, and allows for the rapid convergence of the simulations. Finally, for a given set of $N$, $L$, and $\beta_i$ parameters, 100,000 random runs were carried out to obtain the probability of optimal channel selection as a function of iteration index, \textit{i.e.}, results similar to those presented in Fig.~\ref{fig:prob_N6}. For equal sampling and unequal sampling with $\gamma\in\{-1, -2, -4, -8, -16\}$, the iteration at which the probability of optimal channel selection reaches 0.95 is stored. While the absolute number of iterations required can vary significantly between various system configurations, a normalization has been applied, \textit{i.e.}, the number of iterations required for a given $\gamma$ value is divided by the number of iterations required in the equal allocation strategy. The CDF of this metric is presented in Fig.~\ref{fig:CDF_gamma_change}, where the results are estimated over all $5,200$ simulated system configurations. We observe that for several system configurations, $x$ equals 0.4, \textit{i.e.}, the required probability of the best channel selection is achieved in 40\% of iterations required for the equal allocation algorithm in the case of $\gamma=-8$ or $\gamma=-16$. For 50\% of cases, the best solution is $\gamma=-4$, for which the required detection quality is obtained 30\% faster ($x=0.7$) when compared with the equal allocation strategy. However, for all these $\gamma$ values, there is a small percent of system configurations resulting in a normalized number of iterations required greater than 1 ($x>1$), \textit{i.e.}, equal sampling is a better solution in these cases. The highest probability for this event is obtained for $\gamma=-16$ and equals 14\%. Thus, the recommended solution is to use $\gamma=-2$ since it obtains the lowest probability of being outperformed by the equal samples allocation approach, \textit{i.e.}, occurs in $1.4\%$ of system configurations. At the same time, this approach achieves the required probability of optimal channel selection at least 14\% faster ($x=0.86$) across the 50\% of system configurations. For this performance indicator, the recommended sampling is slightly worse relative to sampling with $\gamma=-4$, $\gamma=-8$ or $\gamma=-16$, but significantly better than when sampling with $\gamma=-1$.            

\begin{table}[]
\caption{Set of simulated $N$ and $L$ configurations}
\centering
\begin{tabular}{|l|l|} 
\hline
\multicolumn{1}{|c|}{$L$} & \multicolumn{1}{c|}{$N$} \\ \hline
3                       & 3,4,5,6,9                  \\ \hline
4                       & 4,5,6,7,8,12               \\ \hline
5                       & 5,6,7,8,9,10,15            \\ \hline
6                       & 6,7,8,9,10,11,12,18        \\ \hline
\end{tabular}
\end{table}

In the following sections, if not stated differently, the unequal sampling algorithm will use $\gamma=-2$.

\section{Bumblebee behavior-based channel sensing and selection in time-varying environment with channel switching cost}
\label{sec:bumblebee}
In Section~\ref{sec:sampl}, we looked at the unequal sampling algorithm for fixed CBR and provided an analytical derivation for the optimal sampling distribution among the sensed frequency channels. In this section, we will evaluate the performance of the considered channel sampling scheme applied in the time-varying CBR scenario. To address the time-varying environment, the knowledge of the prior estimates for the CBR that are stored in a memory will be utilized. Two types of memories will be directly implemented in the bumblebee behavior-based algorithm proposed for the channel selection process. From an implementation perspective, the time-varying CBR is generated using the system-level V2V simulator, described in Section~\ref{sec:sls}, and imported into the analytical simulation framework to evaluate the memory-enabled bumblebee algorithm.

\subsection{Memory-based Bumblebee Algorithm}
In this work, we concentrated on a fully distributed scenario, where each platoon is making its own decision regarding the selection of the best frequency channel for V2V communications. However, at the same time we would like to benefit from the access to the knowledge of prior system states to improve this decision making process. These assumptions are based on bumblebee foraging behavior, where the insects make decisions to select the best flowers while collecting pollen. 
Given the dynamics of these systems, \textit{e.g.}, movement of vehicles, the CBR per channel is expected to change over time. As such, it is reasonable to progressively forget outdated sensing samples, requiring to redefine the CBR estimate for channel $l$, comparing to (\ref{eq_beta_est}), as follows:
\begin{equation}
    \bar{\beta}_{l}^{(i)} =
    \frac{\sum_{j=i-J}^{i} k_l^{(j)}}{\sum_{j=i-J}^{i} N_l^{(j)}} 
\label{eq:mod_beta_est}
\end{equation}
where $J$ is an arbitrarily selected number of sensing iterations.
    
In the original, memoryless bumblebee algorithm \cite{gill2018memory}, the platoon will be making the channel switching decisions based on the instantaneous estimates of the CBR of the candidate channels (\textit{i.e.}, the channels from the set $\mathbb{C}$). However, if the vehicle has access to past CBR values, the performance can be improved by leveraging this memory.
In this paper, both Sliding Window Average (SWA) and Exponentially Weighted Moving Average (EWMA) memory models are evaluated with the number of samples per channel calculated using~(\ref{eq_heuristic}) to smooth out any abrupt transitions in instantaneous sample allocations that may affect the stability of overall implementation. The SWA model results in:
\begin{equation}
\hspace{-89pt}
\bar{\beta}^{(i)\dagger}_l = \dfrac{\sum_{j=0}^{K-1}\bar{\beta}_l^{(i-j)}}{K},
\label{eq:sw}
\end{equation}
being the sliding window average of the candidate channel $l$ over the last $K$ past $\bar{\beta}_l^{(i-j)} (j=0...K-1)$ values observed in $i$-th time interval. 
The other option is to use the EWMA memory model, which we define as:
\begin{align}
\begin{tabular}{c}
$\bar{\beta}_l^{(i)\dagger}$
  \end{tabular}
= \left\{
  \begin{tabular}{l}
 $\bar{\beta}_l^{(0)}$, $i = 0$ \\
   $\alpha \bar{\beta}_l^{(i)} + (1-\alpha)\bar{\beta}_l^{(i-1)\dagger}$, $i > 0$ 
  \end{tabular}
\right.,
 \label{eq:ewma}
\end{align}
where $\alpha$ is the forgetting factor. The larger $\alpha$ value implies that more weight is given to a more recent sample in comparison to the past weighted values. Both memory models are tested with different parameters.

Let us consider the decision to be made by each platoon is either to switch the channel or to stay at the currently selected one. For iteration $i$, we denote the currently selected channel by a given platoon for data transmission as $\hat{l}(i)$. Next, the set of all remaining channels not currently used by this platoon at iteration $i$ is denoted by $\hat{\mathbb{C}}(i) = \mathbb{C}-\hat{l}(i)$. Then, the bumblebee behavior-based channel selection is performed by comparing the CBR with the candidate channel $l^* \in \hat{\mathbb{O}}(i)$, for which the best channel reward is achieved (the lowest CBR, \textit{i.e.}, $l^* = \argmin_{l \in \hat{\mathbb{O}}(i)} \bar{\beta}_l^{(i)\dagger}$). In particular, the following decision making process is conducted:
\begin{align}
\hat{l}(i) =\left\{
  \begin{tabular}{l}
 $l^*$, \hspace{8pt} \text{for} $ 
 \bar{\beta}_{\hat{l}(i-1)}^{(i)\dagger}
 \geq (\bar{\beta}_{l^*}^{(i)\dagger} + \chi)$  \\
   $\hat{l}(i-1)$,\hspace{2pt} \text{for} $ \bar{\beta}_{\hat{l}(i-1)}^{(i)\dagger} < (\bar{\beta}_{l^*}^{(i)\dagger} + \chi)$
  \end{tabular}
\right.,
\label{eq:switchDecision}
\end{align}
where $\bar{\beta}_{\hat{l}(i-1)}^{(i)\dagger}$ is the current channel cost (\textit{i.e.}, observed CBR), $\bar{\beta}_{l^*}^{(i)\dagger}$ is the cost for channel $l^* \in \hat{\mathbb{C}}(i)$, and $\chi$ is the switching cost for the new channel. 
The purpose of the switching cost is to avoid frequent channel changes or even the so-called ``ping-pong'' effect, where the algorithm switches from iteration to iteration between two channels.
Frequent switching between two channels can lead to performance degradation due to the channel-switching hardware-lag, synchronization of the devices, and control signaling.
Thus, for a larger value of $\chi$, this will result in the channel selection being less dynamic. Consequently, the problem of instantaneous and permanent channel switching can be mitigated as changes in the frequency bands will be performed only when new signficant channel opportunities present themselves as it would result from the prior channel assignments. 

Algorithm~\ref{algo} describes the proposed memory-enabled bumblebee VDSA algorithm applied to the V2V communications network, where the non-uniform spectrum sensing decisions are made, and the estimated CBRs are calculated using the number of samples per channel given by the heuristic approach~(\ref{eq_unequal}). The proposed solution consists of two key actions, mainly: (\emph{i}) Sensing Interval, and (\emph{ii}) Transmission interval. With respect to the sensing interval action, the platoon performs spectrum sensing by collecting $N$ samples in the considered frequency band of $L$ channels and facilitates the opportunistic channel access. Initially, the sensing samples are distributed uniformly over all channels allowing for the calculation of $ \bar{\beta}_{l}^{(0)}$. The best channel is selected as $\argmin_{l \in \mathbb{C}} \bar{\beta_{l}^{(0)}}$ and the radio is configured accordingly. In the following iterations, (\ref{eq:switchDecision}) is used to select the best channel with the distribution of sensing samples following the heuristic (\ref{eq_unequal}).
We also assume that during the sensing interval, all the candidate channels can be evaluated by different vehicles, whereas during transmission interval no sensing is performed. Once the channel is selected, the real data transmission is realized.

\begin{algorithm}[!ht]
\caption{Bumblebee-based VDSA Algorithm}
\label{algo}
\SetKwInOut{Input}{Input}\SetKwInOut{Output}{Output}
\Input{$\chi$  - switching cost, $\mathbb{C}$ - set of candidate channels of cardinality $L$, $N$- total number of samples per iteration}
\textbf{Initialize:}
Set $i=0$, and distribute $N$ sensing samples uniformly over $L$ channels. Perform sensing, estimate $\bar{\beta_{l}^{(0)}}$ using (\ref{eq:mod_beta_est}), and assign $\hat{l}(0)=\argmin_{l \in \mathbb{C}} \bar{\beta_{l}^{(0)}}$.

\textbf{Main loop:}\\
\ForEach{ iteration $i$=1,2,3... }
{

Perform $N$ sensing actions distributed among $L$ channels using (\ref{eq_unequal})\;
\ForEach{ candidate channel $l$} {calculate $\bar{\beta}_l^{(i)}$ and store in the memory} 
\If{ SWA memory model is applied }{
Compute $\bar{\beta}_l^{(i)\dagger}$ using (\ref{eq:sw})}
\If{ EWMA  memory model is applied }{
Compute $\bar{\beta}_l^{(i)\dagger}$ using (\ref{eq:ewma})}
Update selected channel $\hat{l}(i)$ according to (\ref{eq:switchDecision})\;
Configure radio with selected channel $\hat{l}(i)$\;
Start V2V transmission;
}
\end{algorithm}


 \subsection{ Computational and Memory Complexity}
{\PS Given how the proposed bumblebee-based VDSA algorithm relies on the analysis of sensing results that can be performed by multiple nodes (vehicles), as well as how it assumes use of memory for storing past values, this algorithm introduces an additional cost to the system in terms of increased computational, memory, and communication overhead. The amount of required additional computations related to CBR estimation should be minimal, as these are performed only by the entity responsible for selecting the frequency band (\textit{e.g.}, the platoon leader) and requires only a limited number of additions and divisions in the calculation of (\ref{eq:mod_beta_est}) and (\ref{eq:sw}), or multiplications and additions when using EWMA for (\ref{eq:ewma}). A slightly more complex scenario is the heuristic allocation of the sensing samples according to (\ref{eq_unequal}) and (\ref{eq_heuristic}) as it requires the calculation of the exponential function. However, the computational overhead is still minimal when compared to the other tasks in a modern wireless transceiver, \textit{e.g.}, decoding.

Taking into account the communication overhead, the information that needs to be distributed from the lead vehicle to the other sensing vehicles is the allocation of sensing samples for each considered band and each vehicle. For example, this can be a vector of $L$ integers denoting in each cell number of samples per each channel to be collected. With respect to the reporting of measurement results, only the number of samples corresponding to a specific channel being busy needs to be communicated (part of the numerator of (\ref{eq:mod_beta_est}), meaning that only single integer vectors of maximum length $L$ are reported by each platoon vehicle. This information can be appended to other messages that are disseminated periodically by platoon vehicles, such as the BSMs.

Finally, the memory cost of the proposed approach is related only to the CBR estimation using (\ref{eq:mod_beta_est}) and averaging using either (\ref{eq:sw}) or (\ref{eq:ewma}). In the worst case, (\ref{eq:mod_beta_est}) requires storage of $2JL$ integers, which corresponds to the number of sensed busy samples $k_l^{(j)}$ and the total number of sensed samples $N_l^{(j)}$ per channel. 
The values of $\bar{\beta}_l^{(i)}$ are subsequently averaged using (\ref{eq:sw}) or (\ref{eq:ewma}) requiring storage of only $K$ or one real value, respectively. Hence, the additional memory utilization is relatively small.}

\subsection{Simulation Results}
\label{sec:results}

\begin{figure}[!ht]
\centering
\includegraphics[width=0.5\textwidth, keepaspectratio=true]{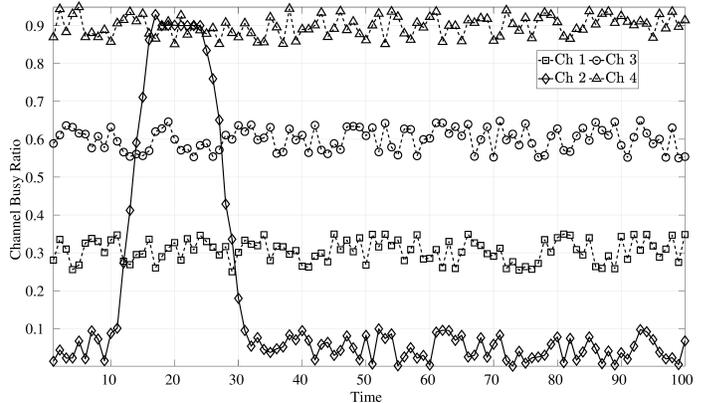}
\caption{Time-varying channel busy ratio generated using the system-level V2V simulator, described in Section~\ref{sec:sls}, with $L~=~4$.}
\label{fig:real_beta}
\end{figure}

We have evaluated the performance of the memory-enabled non-uniform sampling-based bumblebee algorithm using extensive computer simulations. First, the time-varying CBR values were generated in an accurate V2V simulator developed using C++~\cite{vukadinovic20183gpp, sybis2019communication} as illustrated in Fig.~\ref{fig:real_beta}, where the total number of channels is set to $L = 4$. It is observed for the CBR values of the channels under study for the considered period that for three channels the CBR values are slightly oscillating around the specific value (\textit{i.e.} around 0.3 for channel 1, 0.05 for channel 2, 0.6 for channel 3, and 0.9 for channel 4). Such a setup corresponds to the situation where, for example, one channel is frequently occupied by currently ongoing transmissions, whereas the other channels are almost continuously empty. Moreover, a significant change in the CBR is observed in channel~2, which may illustrate in the situation where another platoon utilizing the same band has appeared for some time when overtaking the platoon for which CBRs are shown.  
Additional simulations have been performed using the two different types of memory models employed with the bumblebee algorithm. The total sampling instants $N$ is set to $8$, and $J$ equals 100. For SWA, the $\beta$ values are stored in memory of different lengths, and the bumblebee algorithm makes the switching decision based on the sliding window average given by (\ref{eq:sw}). For the EWMA model, the recent $\beta$ samples are assigned higher weights based on the forgetting factor $\alpha$.

\begin{figure}[!ht]
\centering
\includegraphics[width=0.5\textwidth, keepaspectratio=true]{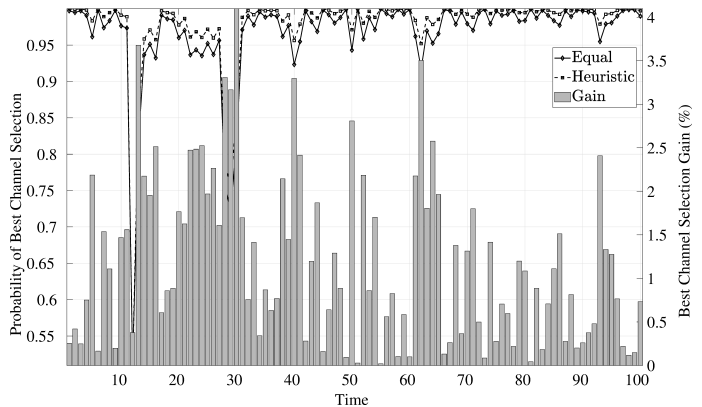}
\caption{Performance comparison for equal sample allocation ($\gamma=0$) and the proposed heuristic approach ($\gamma = -2$). It shows that heuristic approach outperforms equal allocation with no memory. The time from $10$ -- $30$ shows a very high channel utilization where heuristic needs to adapt to the variation but due to no memory the performance is low.}
\label{fig:evsh}
\end{figure}

The performance of the equal allocation approach is compared with the proposed novel heuristic-based unequal allocation scheme using the CBR values generated with the system-level V2V simulator, described in Section~\ref{sec:sls}. Fig.~\ref{fig:evsh} presents a comparison between two schemes: equal samples allocation ($\gamma = 0$) and unequal samples allocation ($\gamma = -2$), with the bar plot representing the gain difference between the two schemes. The gain is defined as the percentage difference between the probabilities of the best channel selection for the two schemes. The gain values (in percentage points) are plotted over the right $y$-axis across simulation time.
Based on this figure, the unequal samples allocation has a consistent gain over equal allocation, which demonstrates the robust capability of the proposed approach. Furthermore, data-driven models can be employed that consider $\gamma$ as a hyper-parameter, and can be tuned based on the environment. Currently, any such approach is outside the scope of this paper and is the subject of future research by the authors.

\begin{figure}[!ht]
    \centering
    \includegraphics[width=0.5\textwidth]{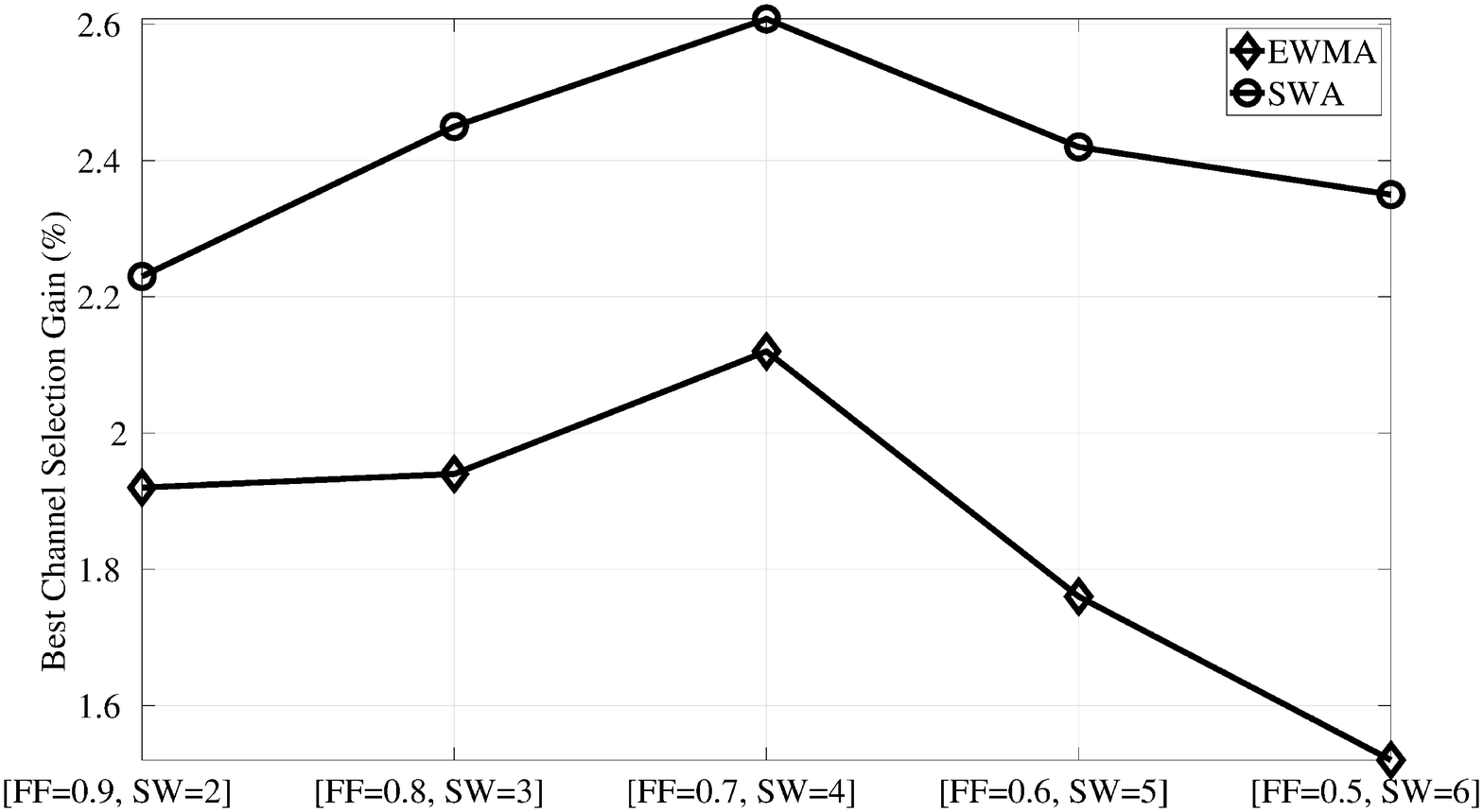}
    \caption{Best Channel Selection gain (\%) is evaluated for different forgetting factors and sliding window memory length. The plot demonstrate the optimal values of both EWMA and SWA memory schemes computed using simulation. }
    \label{fig:comparison}
\end{figure}

The optimal values for the sliding window length $K$ and the forgetting factor $\alpha$ were established using simulations, with the gains compared to the equal allocation shown in Fig.~\ref{fig:comparison}. This result illustrates how the SWA with $K~=~4$ and EWMA with $\alpha~=~0.7$ provide the highest channel selection gain. 

\begin{figure}[!ht]
    \centering
    \includegraphics[width=0.45\textwidth]{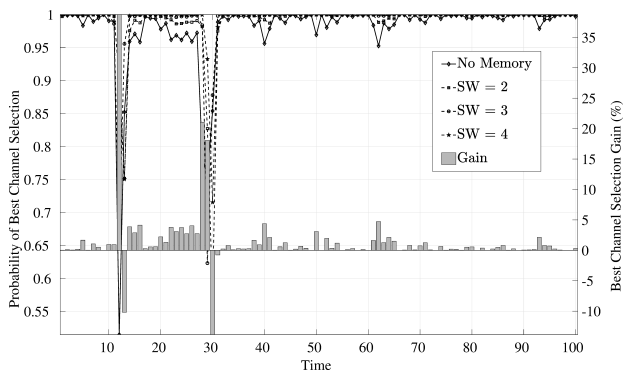}
    \caption{Performance comparison for heuristic approach with $\gamma =-2$ for different memory lengths and equal weight tapering. Memory length $K = 4$ was found to be the best out of the tested configurations. The right axis shows the performance gain over memoryless case.}
    \label{fig:bumblesw}
\end{figure}

Evaluating the impact of the selected memory scheme, Fig.~\ref{fig:bumblesw} describes the advantage of adding memory in the system and exploiting the past values. Three different memory lengths are used, \textit{i.e.}, $K~=~2,~3,~4$, with sliding window average compared against no memory scheme. It is observed that $K~=~4$ provides the highest gain, with increases in the memory length  (\textit{i.e.}, $K > 4$) yielding a degradation in performance as shown in Fig. \ref{fig:comparison}. The best channel selection gain on the right $y$-axis shows the gain for memory of length $K~=~4$ in performance against the memoryless system. 

\begin{figure}[!ht]
    \centering
    \includegraphics[width=0.45\textwidth]{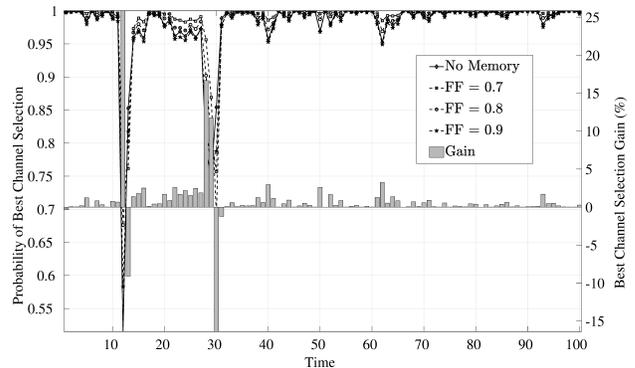}
    \caption{Performance comparison of the heuristic approach when $\gamma = -2$ with different EWMA tapering. $\alpha = 0.7$ appeared to be the best out of the tested configurations. The axis on the right shows the performance gain over memoryless case.}
    \label{fig:bumbleewma}
\end{figure}

In Fig.~\ref{fig:bumbleewma}, we evaluated the performance of the EWMA window for $\alpha~=0.9,~0.8,~0.7$, and compare it against a memoryless scheme. It was observed that $\alpha~=~0.7$ provided the highest gain with decreasing values of $\alpha$ below this threshold leading to diminishing performance. Similar to Fig.~\ref{fig:bumblesw}, the best channel selection gain on the right $y$-axis shows the improvement versus a memoryless system when operating with $\alpha~=~0.7$.

\section{Evaluation in platooning scenario}
\label{sec:sls}
In this section, we present the results of an experimental evaluation of the proposed VDSA algorithm using extensive and sophisticated simulations of the autonomous vehicle platooning scenario, where multiple vehicles follow the platoon leader using the CACC algorithm~\cite{Raj2000}. To facilitate proper CACC operation, platoon vehicles communicate with each other using the IEEE~802.11p protocol~\cite{IEEE80211} in a dynamically selected frequency band, with the platoon leader broadcasting the mobility information to all platoon members, while the other vehicles transmit their position and movement information only to their followers. The bumblebee-based VDSA performance was evaluated using observations of the platoon leader packet successful reception ratio, as the link between the platoon leader and the following vehicles is considered as the weakest link, limiting the performance of the CACC operation~\cite{sybis2019communication}. Furthermore, we studied the latency for the best channel selection (\textit{i.e.}, the delay in switching when the real CBRs facilitate the change of channel) for different channel sampling strategies. {\PS In this evaluation, we considered only two sensing configurations using $\alpha=0.7$: (i)~uniform sampling, and (ii)~memory-based heuristic approach using EWMA.}

\subsection{Simulation Setup}
To evaluate the performance of bumblebee-based VDSA for platooning, we considered a scenario with a 5~km section of a~6-lane highway/motorway, with a single platoon moving {\PS with a velocity of 130~kph} in an outer lane, and with the desired platoon inter-vehicle spacing of 3~m, following the simulation setup introduced in~\cite{Sroka2020b}. Two platooning configurations were evaluated:
\begin{itemize}
    \item A platoon consisting of 4 vehicles (one leader and three followers), thus totaling to an approximate platoon length of 25~m,
    \item A platoon comprising 10 vehicles with an approximate platoon length of 70~m. 
\end{itemize}

Every platoon vehicle transmitted its BSM messages with a 100~ms period in the dynamically selected TVWS band, with each packet comprising of 300~bytes of data~\cite{SAE2735, vukadinovic20183gpp}. There were four 10-MHz frequency bands available to switch between following the VDSA, with center frequencies of 490~MHz, 506~MHz, 522~MHz, and 536~MHz. The bumblebee-based VDSA procedure was applied periodically every 100~ms. Furthermore, each platoon vehicle performed sensing of a single selected frequency band in a $32~{\mu}s$ interval after transmitting its own packet (with a delay between the end of transmission and the start of sensing selected randomly between 1~ms and 5~ms). The sensing results were gathered in a 10~s observation window, corresponding to $J=100$ iterations of the VDSA algorithm, which were used in the calculation of the CBR estimates for the bumblebee-based VDSA. Furthermore, the calculated CBR values were smoothed according to the EWMA rule, defined as in~(\ref{eq:ewma}), with $\alpha=0.7$. Two sensing channel selection methods were considered, with the band allocation for sensing following~(\ref{eq_unequal}) presented in Section~\ref{sec_proposal} for $\gamma=0$ (uniform sampling) and $\gamma=-2$ (non-uniform sampling), respectively. Finally, the channel selection rule proposed in (\ref{eq:switchDecision}) was used assuming the switching cost parameter $\chi=0.1$.\\

Apart from the platoon traveling in the outer lane, there were other vehicles traveling along the other lanes of the highway/motorway, placed randomly following a uniform distribution with an average of 10~vehicles/km/lane. Each non-platoon vehicle was broadcasting 300~bytes BSM messages every 100~ms in a randomly pre-selected frequency band (the band choice remained fixed for the whole simulation run duration), chosen from the set available for the platoon VDSA, with the probabilities of the available channels set as follows: \{0.08, 0.28, 0.16, 0.48\}. Furthermore, there were four roadside units (RSUs) placed along the highway/motorway every 1~km, with the first and the last one transmitting 300~bytes messages every 2~ms in channel~1 (490~MHz), the second transmitting in channel~2 (506~MHz), and the third using channel~3 (522~MHz). These infrastructure transmitters contributed to the significant regional increase in the CBR for selected channels.\\

For both considered platoon configurations, 10 independent simulation runs were performed with the duration of a single simulation run set to 140~s. The main simulation parameters are summarized in Table~\ref{TABLEI}.
\begin{table}[!hbtp]
\centering
\caption{Simulation parameters}
\small
\label{TABLEI}
\begin{tabular}{l|c}
\hline
\textbf{Parameters}  &  \textbf{Values}   \\ \hline
Highway/Motorway length &  5~km \\ \hline
Number of vehicles in platoon &  \{4, 10\} \\ \hline
Inter-vehicle spacing in platoon &  3~m \\ \hline
Messaging periodicity &  100 ms   \\ \hline
BSM message size &  300~bytes        \\ \hline
VDSA frequency bands (channels) &  \{490, 506, 522, 536\}       \\
& MHz\\\hline
VDSA procedure periodicity & 100~ms \\ \hline
Average number of non-platoon vehicles & 10~vehicles/km/lane \\ \hline
Non-platoon vehicles channel probability & \{0.08, 0.28, 0.16, 0.48\}\\ \hline
Number and location of RSUs  & 4 (@ 1, 2, 3, 4~km) \\ \hline
RSU messaging periodicity & 2~ms\\ \hline
Channels used by RSUs &   (1, 2, 3, 1)  \\ \hline
Number of simulations per scenario &    10\\ \hline
Single simulation run duration &     140~s\\ \hline
\end{tabular}
\end{table}
\normalsize

\subsection{Simulation Results for a 4-Vehicle Platoon}
\begin{figure}[!ht]
    \centering
    \includegraphics[width=0.5\textwidth]{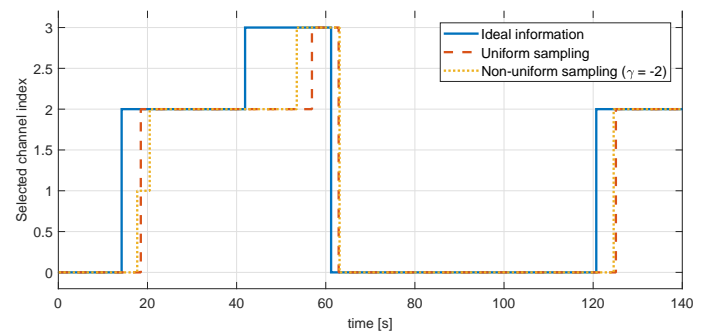}
    \caption{Selected channel (band) index versus time in the case of 4 vehicles platoon.}
    \label{fig:sel_channels_4cars}
\end{figure}

Simulations carried out with a 4-vehicle platoon represents a scenario with a compact entity that performs sensing and VDSA. As the platoon vehicles are closely following one another, the whole platoon length is approximately 25~m, thus the variations in the CBR measured by different platoon vehicles should not significantly impact the results. It is observed in Fig.~\ref{fig:sel_channels_4cars}, where the evolution of the channel selected for transmission in the VDSA procedure is shown for a specified simulation experiment, that for both uniform and non-uniform sampling the correct channel is selected. Note the best available channel, corresponding to the choice with perfect CBR knowledge, is indicated with a solid blue line. However, with the number of samples collected per single VDSA algorithm iteration is relatively small (\textit{e.g.}, 4~samples, which is equal to the number of vehicles in platoon), some latency in channel switching is observed. This latency is smaller in the case of non-uniform sampling, as more sensing slots are allocated to the more promising channels here compared to the case of uniform sampling. Therefore, using non-uniform sampling enables the VDSA algorithm to switch faster to a better channel when the experienced CBRs change.

\begin{figure}[!ht]
    \centering
    \includegraphics[width=0.5\textwidth]{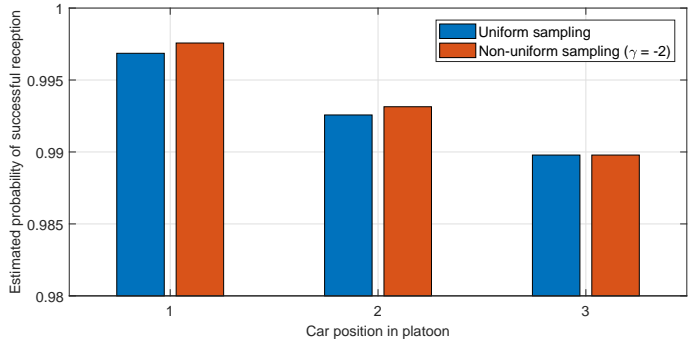}
    \caption{Estimates of the probability of successful reception of leader's packets versus vehicle position in a 4-vehicle platoon.}
    \label{fig:reception_4cars}
\end{figure}

\begin{figure}[!ht]
    \centering
    \includegraphics[width=0.5\textwidth]{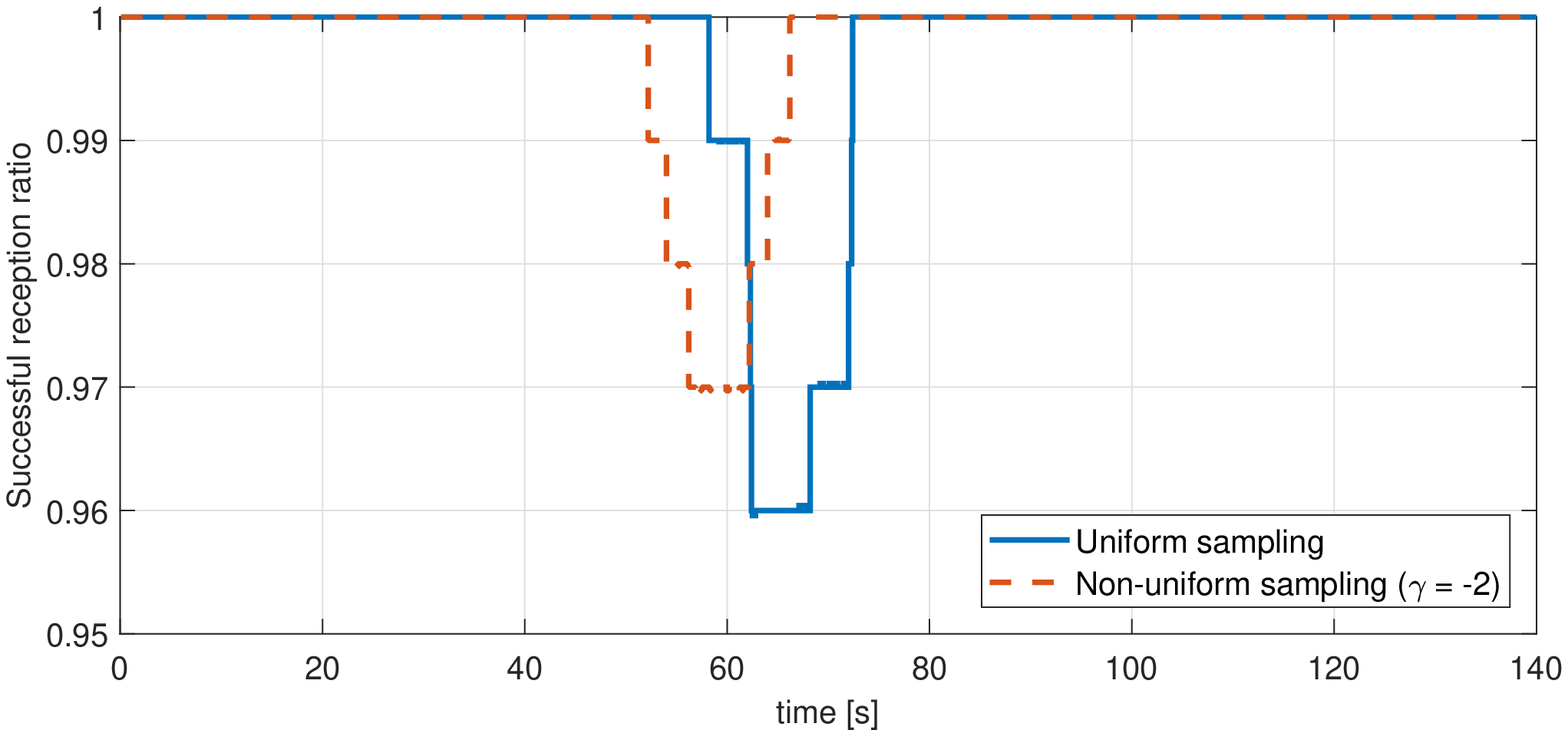}
    \caption{Example of time evolution of the successful reception ratio of leader's packets observed in a window of 10~s for vehicle~2 in a 4-vehicle platoon.}
    \label{fig:evolution_4cars}
\end{figure}

The ability to quickly detect the channel with the lowest CBR using non-uniform sampling impacts the performance of intra-platoon communications. Referring to Fig.~\ref{fig:reception_4cars}, we observe the estimated probability of successful reception of leader packets versus vehicle position in platoon averaged over all simulation runs. In this figure, there exists a slightly higher probability that vehicles~1 and 2 with non-uniform sampling can switch faster to a better band when the CBR values change. This phenomenon is further highlighted in Fig.~\ref{fig:evolution_4cars}, which presents the evolution of the successful reception of leader packets (measured in a moving 10~s observation window) versus the time for 
a selected simulation experiment. The reception rate drops in both considered cases around $t=60~s$ due to the increasing occupancy of the currently used channel. It should be noted that with non-uniform sampling the switch is performed faster, thus yielding a in smaller drop in successful reception ratio.

\begin{figure}[!ht]
    \centering
    \includegraphics[width=0.5\textwidth]{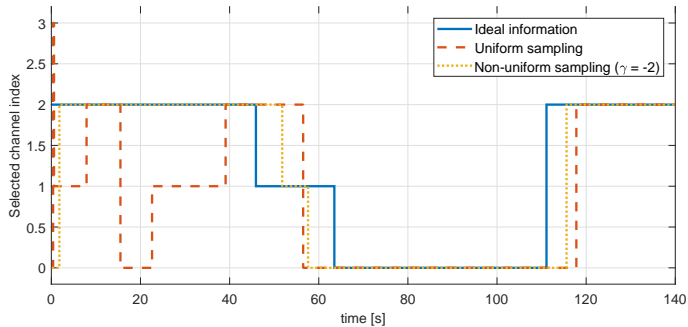}
    \caption{Selected channel (band) index versus time in the case of a 10-vehicle platoon.}
    \label{fig:sel_channels_10cars}
\end{figure}

\subsection{Simulation Results for a 10-Vehicle Platoon}
\begin{figure}[!ht]
    \centering
    \includegraphics[width=0.5\textwidth]{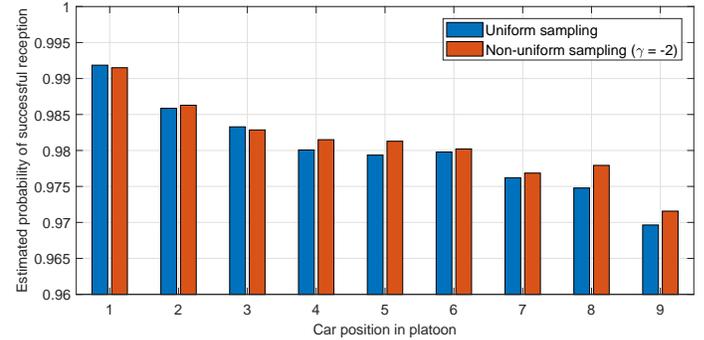}
    \caption{Estimates of the probability of successful reception of leader's packets versus vehicle position in a 10-vehicle platoon.}
    \label{fig:reception_10cars}
\end{figure}

The scenario with a 10-vehicle platoon represents a more complicated situation, where the platoon length is significantly larger and might exceed 70~m. In such a case, the sensing results obtained for different platoon vehicles might differ significantly, thus affecting the ability to select the lowest CBR channel in the VDSA procedure. This is clearly visible in Fig.~\ref{fig:reception_10cars}, which presents the evolution of the channel selected for transmission in the VDSA procedure in a selected simulation run, where for the uniform sampling case the VDSA algorithm switches frequently between bands, and it is not able to find the best one (represented with a solid blue line) for the initial 60~s. This results from a very limited number of samples that are collected, with only a single sample per vehicle collected in each iteration, thus causing the CBR values to be dependent on the location of a vehicle measuring the channel of interest. With non-uniform sampling, more sensing slots are allocated to the potentially more suitable channels, thus resulting in a higher level of accuracy of the CBR estimation. Thus, the selected channel with non-uniform sampling follows closely the reference (with perfect CBR knowledge) best channel, with only some latency introduced due to the averaging of results in a 10~s observation window.

The improved ability to find the lowest CBR channel with the non-uniform sampling approach is also reflected in the estimates of the probability of successful reception of the leader packets, as presented in Fig.~\ref{fig:reception_10cars}, with the results averaged over 10~simulation runs. Based on these results, we observe that with non-uniform sampling the VDSA algorithm is is capable of reducing the losses of the transmitted packets, especially for the vehicles at the end of the platoon, where the reception rate is usually lower due to higher channel attenuation.

\begin{figure}[!ht]
    \centering
    \includegraphics[width=0.5\textwidth]{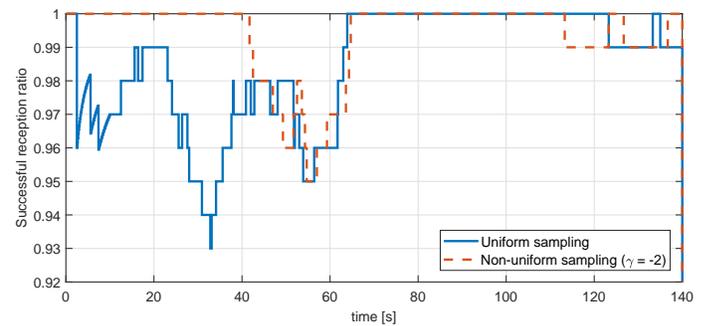}
    \caption{Example of time evolution of the successful reception ratio of leader's packets observed in a window of 10~s for vehicle~9 in a 10-vehicle platoon.}
    \label{fig:evolution_10cars}
\end{figure}

The reduced ability for selecting a proper channel with uniform sampling can be also observed in Fig.~\ref{fig:evolution_10cars}, which shows the evolution of the successful reception of the leader packets (measured within a moving 10~s observation window) versus the time for a selected simulation run. With uniform sampling, the VDSA algorithm is unable to find the channel guaranteeing sufficient quality of transmission for the initial 60~s, thus resulting in a temporary packet error rate over 5\%, which can be unacceptable with autonomous platooning. With non-uniform sampling, only a short term and less significant drop in successful reception ratio is observed around $t=60~s$ that is the result of rapid changes in CBR values of the three channels.

Overall, for a platooning scenario employing the bumblebee-based VDSA algorithm, the intra-platoon communication channels can be selected in a dynamic manner.  However, the implementation needs to be provided with a fairly accurate set of CBR values. These can be reliably obtained by providing a sufficient number of sensing samples for the prospective frequency bands. With non-uniform sampling these are secured faster for the channels of interest (\textit{i.e.}, the channels with low CBR) than with uniform sensing slots allocation.

\section{Conclusion}
\label{sec:conclusion}
In this work, we presented the framework for a memory-enabled bumblebee foraging algorithm for vehicular platoon communications. The optimized unequal sampling allocation heuristic is proposed to estimate the CBR  $\beta$ with sufficiently high accuracy. Based on the results obtained, the unequal sampling instant allocation approach outperforms the equal sampling allocation scheme with the proposed sub-optimal allocation heuristic approach. We have also implemented two memory models that are integrated with the bumblebee foraging algorithm to leverage available memory, which boosts the probability of the best channel selection. Sliding window average and exponentially weighted moving average memory schemes are employed and their performance is compared against the memoryless model. Using the sliding window average memory scheme, different memory lengths are utilized, and, similarly, different forgetting factors are used for the exponentially weighted moving average scheme. The simulation results show the bumblebee algorithm with unequal sampling instants allocation heuristic approach provides higher accuracy compared to the equal allocation scheme, especially in a scenario where the sensing resources are scarce.

For future work, the authors will explore a holistic memory model that can adapt to different time-varying conditions, which can be tuned by employing machine learning. A new heuristic approach can also be studied where $\gamma$ can be dynamically adjusted based on the sampling instants and CBR variations of the channels. Finally, we would like to evaluate the proposed algorithm in over-the-air testing using software-defined radio (SDR) testbed technology.

\section*{Acknowledgment}
The authors would like to acknowledge the generous support of this research by the US National Science Foundation via Award Number 1547296 and the National Science Centre of Poland via Project Number 2018/29/B/ST7/01241.

\bibliographystyle{IEEEtran}
\bibliography{biblio}

\begin{thebibliography}{10}
\providecommand{\url}[1]{#1}
\csname url@samestyle\endcsname
\providecommand{\newblock}{\relax}
\providecommand{\bibinfo}[2]{#2}
\providecommand{\BIBentrySTDinterwordspacing}{\spaceskip=0pt\relax}
\providecommand{\BIBentryALTinterwordstretchfactor}{4}
\providecommand{\BIBentryALTinterwordspacing}{\spaceskip=\fontdimen2\font plus
\BIBentryALTinterwordstretchfactor\fontdimen3\font minus
  \fontdimen4\font\relax}
\providecommand{\BIBforeignlanguage}[2]{{%
\expandafter\ifx\csname l@#1\endcsname\relax
\typeout{** WARNING: IEEEtran.bst: No hyphenation pattern has been}%
\typeout{** loaded for the language `#1'. Using the pattern for}%
\typeout{** the default language instead.}%
\else
\language=\csname l@#1\endcsname
\fi
#2}}
\providecommand{\BIBdecl}{\relax}
\BIBdecl

\bibitem{park2011integrated}
H.~Park, A.~Miloslavov, J.~Lee, M.~Veeraraghavan, B.~Park, and B.~L. Smith,
  ``Integrated traffic--communication simulation evaluation environment for
  intellidrive applications using {SAE} {J2735} message sets,''
  \emph{Transportation research record}, vol. 2243, no.~1, pp. 117--126, 2011.

\bibitem{Shi2014}
L.~Shi and K.~W. Sung, ``Spectrum requirement for vehicle-to-vehicle
  communication for traffic safety,'' in \emph{2014 IEEE 79th Vehicular
  Technology Conference (VTC Spring)}, 2014, pp. 1--5.

\bibitem{chen2011feasibility}
S.~Chen, A.~M. Wyglinski, S.~Pagadarai, R.~Vuyyuru, and O.~Altintas,
  ``Feasibility analysis of vehicular dynamic spectrum access via queueing
  theory model,'' \emph{IEEE Communications Magazine}, vol.~49, no.~11, pp.
  156--163, 2011.

\bibitem{pagadarai2009characterization}
S.~Pagadarai, A.~M. Wyglinski, and R.~Vuyyuru, ``Characterization of vacant
  {UHF} {TV} channels for vehicular dynamic spectrum access,'' in \emph{2009
  IEEE Vehicular Networking Conference (VNC)}.\hskip 1em plus 0.5em minus
  0.4em\relax IEEE, 2009, pp. 1--8.

\bibitem{mishra2010much}
S.~Mishra and A.~Sahai, ``How much white space has the {FCC} opened up?''
  \emph{IEEE Communication Letters}, 2010.

\bibitem{kryszkiewicz2021dynamic}
P.~Kryszkiewicz, H.~Kokkinen, J.~Ojaniemi, and D.~Sonoiya, ``Dynamic spectrum
  access in terrestrial {TV} band: assessment of prospects in kenya,''
  \emph{Telecommunication Systems}, pp. 1--11, 2021.

\bibitem{kenney2011dedicated}
J.~B. Kenney, ``Dedicated short-range communications ({DSRC}) standards in the
  united states,'' \emph{Proceedings of the IEEE}, vol.~99, no.~7, pp.
  1162--1182, 2011.

\bibitem{chen2020vision}
S.~Chen, J.~Hu, Y.~Shi, L.~Zhao, and W.~Li, ``A vision of {C-V2X}:
  technologies, field testing, and challenges with chinese development,''
  \emph{IEEE Internet of Things Journal}, vol.~7, no.~5, pp. 3872--3881, 2020.

\bibitem{8746562}
V.~Mannoni, V.~Berg, S.~Sesia, and E.~Perraud, ``A comparison of the {V2X}
  communication systems: {ITS-G5} and {C-V2X},'' in \emph{2019 IEEE 89th
  Vehicular Technology Conference (VTC2019-Spring)}, 2019, pp. 1--5.

\bibitem{Liu2021}
X.~Liu, C.~Sun, M.~Zhou, B.~Lin, and Y.~Lim, ``Reinforcement learning based
  dynamic spectrum access in cognitive internet of vehicles,'' \emph{China
  Communications}, vol.~18, no.~7, pp. 58--68, 2021.

\bibitem{Arteaga2019}
A.~Arteaga, S.~Céspedes, and C.~Azurdia-Meza, ``Vehicular communications over
  tv white spaces in the presence of secondary users,'' \emph{IEEE Access},
  vol.~7, pp. 53\,496--53\,508, 2019.

\bibitem{vukadinovic20183gpp}
V.~Vukadinovic, K.~Bakowski, P.~Marsch, I.~D. Garcia, H.~Xu, M.~Sybis,
  P.~Sroka, K.~Wesolowski, D.~Lister, and I.~Thibault, ``{3GPP} {C-V2X} and
  {IEEE} 802.11p for vehicle-to-vehicle communications in highway platooning
  scenarios,'' \emph{Ad Hoc Networks}, vol.~74, pp. 17--29, 2018.

\bibitem{sybis2018context}
M.~Sybis, P.~Kryszkiewicz, and P.~Sroka, ``On the context-aware, dynamic
  spectrum access for robust intraplatoon communications,'' \emph{Mobile
  Information Systems}, 2018.

\bibitem{Wei_2013}
Z.~{Wei}, Q.~{Zhang}, Z.~{Feng}, W.~{Li}, and T.~A. {Gulliver}, ``On the
  construction of radio environment maps for cognitive radio networks,'' in
  \emph{2013 IEEE Wireless Communications and Networking Conference (WCNC)},
  April 2013, pp. 4504--4509.

\bibitem{urkowitz1967energy}
H.~Urkowitz, ``Energy detection of unknown deterministic signals,''
  \emph{Proceedings of the IEEE}, vol.~55, no.~4, pp. 523--531, 1967.

\bibitem{Bazzi2019}
A.~Bazzi, ``Congestion control mechanisms in {IEEE} 802.11p and sidelink
  {C-V2X},'' in \emph{2019 53rd Asilomar Conference on Signals, Systems, and
  Computers}, 2019, pp. 1125--1130.

\bibitem{Lehtomaki2015_CBR_estimation}
J.~J. {Lehtomäki}, M.~{López-Benítez}, K.~{Umebayashi}, and M.~{Juntti},
  ``Improved channel occupancy rate estimation,'' \emph{IEEE Transactions on
  Communications}, vol.~63, no.~3, pp. 643--654, 2015.

\bibitem{gill2018memory}
K.~S. Gill, B.~Aygun, K.~N. Heath, R.~J. Gegear, E.~F. Ryder, and A.~M.
  Wyglinski, ``Memory matters: Bumblebee behavioral models for
  vehicle-to-vehicle communications,'' \emph{IEEE Access}, vol.~6, pp.
  25\,437--25\,447, 2018.

\bibitem{cheng2011cognitive}
X.~Cheng and M.~Jiang, ``Cognitive radio spectrum assignment based on
  artificial bee colony algorithm,'' in \emph{2011 IEEE 13th International
  Conference on Communication Technology}.\hskip 1em plus 0.5em minus
  0.4em\relax IEEE, 2011, pp. 161--164.

\bibitem{li2013adaptive}
G.~Li and L.~Boukhatem, ``Adaptive vehicular routing protocol based on ant
  colony optimization,'' in \emph{Proceeding of the tenth ACM international
  workshop on vehicular inter-networking, systems, and applications}, 2013, pp.
  95--98.

\bibitem{boban2014geometry}
M.~Boban, J.~Barros, and O.~K. Tonguz, ``Geometry-based vehicle-to-vehicle
  channel modeling for large-scale simulation,'' \emph{IEEE Transactions on
  Vehicular Technology}, vol.~63, no.~9, pp. 4146--4164, 2014.

\bibitem{gemv2}
\BIBentryALTinterwordspacing
{GEMV2}. [Online]. Available: \url{http://vehicle2x.net/}
\BIBentrySTDinterwordspacing

\bibitem{gill2018capacity}
K.~Gill, K.~N. Heath, R.~J. Gegear, E.~F. Ryder, and A.~M. Wyglinski, ``On the
  capacity bounds for bumblebee-inspired connected vehicle networks via queuing
  theory,'' in \emph{2018 IEEE 87th vehicular technology conference (VTC
  spring)}.\hskip 1em plus 0.5em minus 0.4em\relax IEEE, 2018, pp. 1--6.

\bibitem{gill2018experimental}
K.~S. Gill, K.~McClintick, N.~Kanthasamy, J.~Tolbert, D.~Nguyen, S.~Nguyen,
  G.~Wernsing, V.~Moore, I.~Gelman, A.~O'Neil, N.~Schubert, C.~Coogan,
  K.~Murdy, B.~Mahan, S.~Halama, K.~N. Heath, E.~F. Ryder, R.~J. Gegear, and
  A.~M. Wyglinski, ``Experimental test-bed for bumblebee-inspired channel
  selection in an ad-hoc network,'' in \emph{2018 IEEE 88th Vehicular
  Technology Conference (VTC-Fall)}, 2018, pp. 1--5.

\bibitem{pluto}
\BIBentryALTinterwordspacing
{Pluto-{SDR}}. [Online]. Available:
  \url{https://www.analog.com/en/design-center/evaluation-hardware-and-software/evaluation-boards-kits/adalm-pluto/}
\BIBentrySTDinterwordspacing

\bibitem{sybis2019communication}
M.~Sybis, V.~Vukadinovic, M.~Rodziewicz, P.~Sroka, A.~Langowski, K.~Lenarska,
  and K.~Weso{\l}owski, ``Communication aspects of a modified cooperative
  adaptive cruise control algorithm,'' \emph{IEEE Transactions on Intelligent
  Transportation Systems}, vol.~20, no.~12, pp. 4513--4523, 2019.

\bibitem{IEEE80211}
``{IEEE} {S}tandard for {I}nformation technology--{T}elecommunications and
  information exchange between systems local and metropolitan area
  networks--{S}pecific requirements -- {P}art 11: {W}ireless {LAN} {M}edium
  {A}ccess {C}ontrol ({MAC}) and {P}hysical {L}ayer ({PHY}) {S}pec.'' Dec.
  2016.

\bibitem{Raj2000}
R.~Rajamani, H.-S. Tan, B.~K. Law, and W.-B. Zhang, ``Demonstration of
  integrated longitudinal and lateral control for the operation of automated
  vehicles in platoons,'' \emph{IEEE Transactions on Control Systems
  Technology}, vol.~8, no.~4, pp. 695--708, 2000.

\bibitem{Sroka2020b}
P.~{Sroka} \emph{et~al.}, ``Distributed {V}ehicular {D}ynamic {S}pectrum
  {A}ccess for {P}latooning environments,'' in \emph{2020 IEEE 91st Vehicular
  Technology Conf. (VTC2020-Spring)}, 2020, pp. 1--5.

\bibitem{SAE2735}
``{SAE} {S}tandard {J2735}: {D}edicated {S}hort {R}ange {C}ommunications
  ({DSRC}) {M}essage {S}et {D}ictionary,'' Jan. 2016.

\end{thebibliography}

\end{document}